\documentclass[twocolumn,tighten,times]{aastex62}

\hypersetup{linkcolor=black,citecolor=black,filecolor=cyan,urlcolor=magenta,allcolors=blue}

\usepackage{amsmath}
\usepackage{longtable}
\usepackage{graphicx}
\usepackage{rotating}
\usepackage{tikz}
\usepackage{float}
\usepackage{lipsum}
\usepackage[para]{threeparttable}
\usepackage{isotope}
\usepackage{supertabular}

\shorttitle{SN 2010kd, a Slow-Decaying SLSN~I}
\shortauthors{Kumar et al.}

\begin{document}

\title{SN 2010kd: Photometric and Spectroscopic Analysis of a Slow-Decaying Superluminous Supernova}

\correspondingauthor{Amit Kumar, S. B. Pandey}
\email{amit@aries.res.in, shashi@aries.res.in}

\author{Amit Kumar}
\affil{Aryabhatta Research Institute of Observational Sciences, Manora Peak, Nainital, 263 002, India}
\affil{School of Studies in Physics and Astrophysics, Pandit Ravishankar Shukla University, Chattisgarh 492 010, India}

\author{Shashi Bhushan Pandey}
\affil{Aryabhatta Research Institute of Observational Sciences, Manora Peak, Nainital, 263 002, India}

\author{Reka Konyves-Toth}
\affil{Konkoly Observatory, Research Center for Astronomy and Earth Sciences, Konkoly Thege M. ut 15-17, Budapest 1121, Hungary}

\author{Ryan Staten}
\affil{Department of Physics, Southern Methodist University, 3215 Daniel Ave., Dallas, TX 75205, USA}

\author{Jozsef Vinko}
\affil{Konkoly Observatory, Research Center for Astronomy and Earth Sciences,
Konkoly Thege M. ut 15-17, Budapest 1121, Hungary}
\affil{Department of Optics and Quantum Electronics, University of Szeged, Dom ter 9, Szeged 6720, Hungary}
\affil{Department of Astronomy, University of Texas, Austin, TX 79712, USA}

\author{J. Craig Wheeler}
\affil{Department of Astronomy, University of Texas, Austin, TX 79712, USA}

\author{WeiKang Zheng}
\affil{Department of Astronomy, University of California, Berkeley, CA 94720-3411, USA}

\author{Alexei V. Filippenko}
\affil{Department of Astronomy, University of California, Berkeley, CA 94720-3411, USA}
\affil{Miller Senior Fellow, Miller Institute for Basic Research in Science, University of California, Berleley, CA 94720, USA}

\author{Robert Kehoe}
\affil{Department of Physics, Southern Methodist University, 3215 Daniel Ave., Dallas, TX 75205, USA}

\author{Robert Quimby}
\affil{Department of Astronomy/Mount Laguna Observatory, San Diego State University, 5500 Campanile Drive,
San Diego, CA 92812-1221, USA}
\affil{Kavli IPMU (WPI), UTIAS, The University of Tokyo, Kashiwa, Chiba 277-8583, Japan}

\author{Yuan Fang}
\affil{Department of Physics, University of Michigan, 450 Church Street, Ann Arbor, MI 48109-1040, USA}

\author{Carl Akerlof}	
\affil{Department of Physics, University of Michigan, 450 Church Street, Ann Arbor, MI 48109-1040, USA}

\author{Tim A. McKay}
\affil{Department of Physics, University of Michigan, 450 Church Street, Ann Arbor, MI 48109-1040, USA}

\author{Emmanouil Chatzopoulos}
\affil{Department of Astronomy, University of Texas, Austin, TX 79712, USA}
\affil{Department of Physics \& Astronomy, Louisiana State University, Baton Rouge, LA 70803, USA}

\author{Benjamin P. Thomas}
\affil{Department of Astronomy, University of Texas at Austin, Austin, TX, USA}

\author{Govinda Dhungana}
\affil{Department of Physics, Southern Methodist University, 3215 Daniel Ave., Dallas, TX 75205, USA}

\author{Amar Aryan}
\affil{Aryabhatta Research Institute of Observational Sciences, Manora Peak, Nainital, 263 002, India}
\affil{Department of Physics, Deen Dayal Upadhyaya Gorakhpur University, Gorakhpur 273009, India}

\author{Raya Dastidar}
\affil{Aryabhatta Research Institute of Observational Sciences, Manora Peak, Nainital, 263 002, India}
\affil{Department of Physics \& Astrophysics, University of Delhi, Delhi-110 007, India}

\author{Anjasha Gangopadhyay}
\affil{Aryabhatta Research Institute of Observational Sciences, Manora Peak, Nainital, 263 002, India}
\affil{School of Studies in Physics and Astrophysics, Pandit Ravishankar Shukla University, Chattisgarh 492 010, India}

\author{Rahul Gupta}
\affil{Aryabhatta Research Institute of Observational Sciences, Manora Peak, Nainital, 263 002, India}
\affil{Department of Physics, Deen Dayal Upadhyaya Gorakhpur University, Gorakhpur 273009, India}

\author{Kuntal Misra}
\affil{Aryabhatta Research Institute of Observational Sciences, Manora Peak, Nainital, 263 002, India}

\author{Brajesh Kumar}
\affil{Aryabhatta Research Institute of Observational Sciences, Manora Peak, Nainital, 263 002, India}

\author{Nameeta Brahme}
\affil{School of Studies in Physics and Astrophysics, Pandit Ravishankar Shukla University, Chattisgarh 492 010, India}

\author{David Buckley}
\affil{South African Astronomical Observatory, P.O. Box 9, Observatory 7935, Cape Town, South Africa}

\begin{abstract}
This paper presents data and analysis of SN 2010kd, a low-redshift ($z = 0.101$) H-deficient superluminous supernova (SLSN), based on ultraviolet/optical photometry and optical spectroscopy spanning between $-$28 and +194 days relative to \textit{B}-band maximum light. The \textit{B}-band light-curve comparison of SN 2010kd with a subset of well-studied SLSNe~I at comparable redshifts indicates that it is a slow-decaying PTF12dam-like SLSN. Analytical light-curve modeling using the {\tt Minim} code suggests that the bolometric light-curve of SN 2010kd favors circumstellar matter interaction for the powering mechanism. {\tt SYNAPPS} modeling of the early-phase spectra does not identify broad H or He lines, whereas the photospheric-phase spectra are dominated by O~I, O~II, C~II, C~IV and Si~II, in particular the presence of both low- and high-velocity components of O~II and Si~II lines. The nebular-phase spectra of SN 2010kd are dominated by O~I and Ca~II emission lines similar to those seen in other SLSNe~I. The line velocities in SN 2010kd exhibit flatter evolution curves similar to SN 2015bn but with comparatively higher values. SN 2010kd shows a higher single-zone local thermodynamic equilibrium temperature in comparison to PTF12dam and SN 2015bn, and it has an upper O~I ejected mass limit of $\sim 10~M_\odot$. The host of SN 2010kd is a dwarf galaxy with a high star-formation rate ($\sim 0.18 \pm 0.04~M_\odot$ yr$^{-1}$) and extreme emission lines.
\end{abstract}


\section{Introduction} \label{sec:int}
Superluminous supernovae (SLSNe) are the most luminous SNe, having a mean absolute magnitude of $\sim -21.7$ mag \citep{Gal-Yam2012,Quimby2013,Nicholl2016a,DeCia2018,Angus2019,Inserra2019}. SLSNe are very rare, comprising only $\sim 0.01$\% of the normal population of core-collapse SNe \citep[CCSNe;][]{Quimby2013,McCrum2015,Prajs2017}. The era of SLSN studies started with the discovery of SN 2005ap \citep{Quimby2007} by the Robotic Optical Transient Search Experiment \citep[ROTSE;][]{Akerlof2003}. Later, new sky surveys came into existence and began discovering SLSNe at a rate of $\sim 3$ per month; see, for example, the Palomar Transient Factory \citep[PTF;][]{Quimby2011}, the Pan-STARRS1 \citep[PS1;][]{Chornock2013}, and the Zwicky Transient Facility \citep[ZTF;][]{Lunnan2018}. Being highly luminous, SLSNe can be used to reveal and understand the last evolutionary phases of very massive stars and to probe the high-redshift universe. The host galaxies of SLSNe are usually faint dwarf galaxies having low metallicity, with extremely strong emission lines suggesting enhanced star-forming activity \citep{Lunnan2014,Leloudas2015,Perley2016,Chen2017b,Schulze2018}.

Like normal SNe, SLSNe are spectroscopically divided into two categories: hydrogen-poor (SLSNe~I) and hydrogen-rich \citep[SLSNe~II;][]{Gal-Yam2012}. SLSNe~I are thought to be a subtype of SNe~Ic, because after a few weeks, they exhibit spectral features similar to those of SNe~Ic \citep{Pastorello2010,Inserra2013}. SLSNe~I are broadly classified into fast- and slow-decaying categories \citep{Quimby2011,Quimby2018,Inserra2017,Inserra2019}; however, they seem to have a wider range of postpeak decay rates with no significant gap \citep{Nicholl2015a,DeCia2018}.

SLSNe are among one of the least understood SNe because not only are their underlying progenitors unclear, but their extremely high peak luminosity is also unexplained using conventional SN power-source models \citep{Gal-Yam2012}. Various studies have suggested different power sources for SLSNe~I \citep{Moriya2018,Wang2019}. To explain the required high peak luminosity of SLSNe~I with the primary power source being the radioactive decay (RD) of \isotope[56]{Ni} $\rightarrow$ \isotope[56]{Co} $\rightarrow$ \isotope[56]{Fe}, the synthesized \isotope[56]{Ni} mass ($M_{\rm Ni}$) should be $\gtrsim 5\,M_\odot$ \citep{Gal-Yam2012}; however, the synthesized $M_{\rm Ni}$ in CCSNe cannot exceed $\sim 4\,M_\odot$ \citep{Umeda2008}. Moreover, late-time observations of SLSNe~I suggest that $M_{\rm Ni} \lesssim 1\,M_\odot$ for many SLSNe~I \citep{Pastorello2010,Chen2013}, inconsistent with the $M_{\rm Ni}$ estimated by the RD model. Also, the median of the observed ejecta masses ($M_{\rm ej}$) calculated using a sample of SLSNe~I is $\sim 6\,M_\odot$ \citep{Nicholl2015a}. This value is closer to (or sometimes even lower than) the $M_{\rm Ni}$ obtained from the RD model, which makes this model unphysical. The pair-instability SN (PISN) model was also suggested as a power mechanism for some slow-decaying SLSNe~I (e.g., SN 2007bi; \citealt{Gal-Yam2009}), but later it was found that the bluer color and the fast-rising rate of slow-decaying SLSNe~I contradict the PISN model \citep{Kasen2011,Dessart2012,Nicholl2013,Jerkstrand2017}.

Spin-down millisecond magnetars having magnetic fields of a few $\times 10^{14}$ G are also considered as possible power sources for many SLSNe~I \citep{Kasen2010,Woosley2010,Inserra2013,Dessart2019}. In addition, a subset of SNe~Ib/c and some gamma-ray bursts (GRBs) may have magnetars as central power sources \citep{Wheeler2000,Metzger2011}. However, contrary to this, many SLSNe~I have prepeak bumps (e.g., LSQ14bdq; \citealt{Nicholl2015b}) and postpeak undulations (e.g., SN 2015bn; \citealt{Nicholl2015a}) in their light-curves that cannot be explained well using a spin-down magnetar (MAG) model.

The circumstellar matter interaction (CSMI) is also suggested as a possible powering mechanism for many SLSNe~I \citep{Chevalier2011}. Recently, \cite{Wheeler2017} found that SN 2017egm showed an irregular light-curve with a convex rise to peak, which is inexplicable by either the MAG or RD models.
Both of these models demand a concave downward rise and decline from the peak, so SN 2017egm proved to be powered by the CSMI. Conversely, SLSNe~I do not show any signatures of CSM interaction in their spectra, except for iPTF13ehe \citep{Yan2015}, iPTF15esb, and iPTF16bad \citep{Yan2017a}. In many cases, light-curve fitting of SLSNe~I requires a combination of the above models; we call these ``HYBRID'' models based on \cite{Chatzopoulos2012}. For example, the primary power source may be a central engine, but a small amount of CSMI might also play role \citep{Chen2017a,Inserra2017}.

In this paper, we present the analysis of the optical photometric and spectroscopic data of the nearby SN 2010kd. 
For the photometric study, early-time ROTSE-IIIb discovery data along with multiband 
{\it Swift}-UVOT data and other published optical photometry taken from \cite{Roy2012} in the Johnson \textit{UBV} and Cousins \textit{RI} filters are used. The spectral data for SN 2010kd were obtained using the 9.2~m Hobby-Eberly Telescope (HET-9.2m) and the Keck-I 10~m telescope (Keck-10m).
 
Section~\ref{sec:dis} discusses the multiband light-curve analysis and analytical fitting to the bolometric light-curve using {\tt MINIM}. The optical light-curve comparison of SN 2010kd with some well-studied SLSNe~I at comparable redshift is presented in Section~\ref{sec:ligcomp}. Section~\ref{sec:spectroscopy} describes the spectroscopic study of SN 2010kd along with the {\tt SYNAPPS} spectral modeling of the photospheric-phase spectra. In Section~\ref{sec:comwithother}, we compare spectra of the SN 2010kd with those of some slow- and fast-decaying SLSNe~I and several SNe~Ic-BL. Section~\ref{sec:velocomp} presents a velocity comparison of various identified elements of SN 2010kd with a set of well-studied SLSNe~I at comparable redshift. In Section~\ref{sec:linelum}, we determine various physical parameters of line luminosities during the nebular-phase, their ratios, and comparison with a set of SLSNe~I. Constraints on the host-galaxy properties are discussed in Section~\ref{sec:host}. We summarize results in Section~\ref{sec:CONclusion}. Throughout this paper, the temporal phase is referred with respect to the \textit{B}-band maximum light and the magnitudes are expressed in Vega. The phase before \textit{B}-band maximum is considered to be the hot photospheric-phase; after maximum, it is termed as the cool photospheric-phase and after +90 days designated as the nebular-phase \citep{Gal-Yam2019}.

\begin{figure*}[ht!]
\includegraphics[angle=0,scale=0.73]{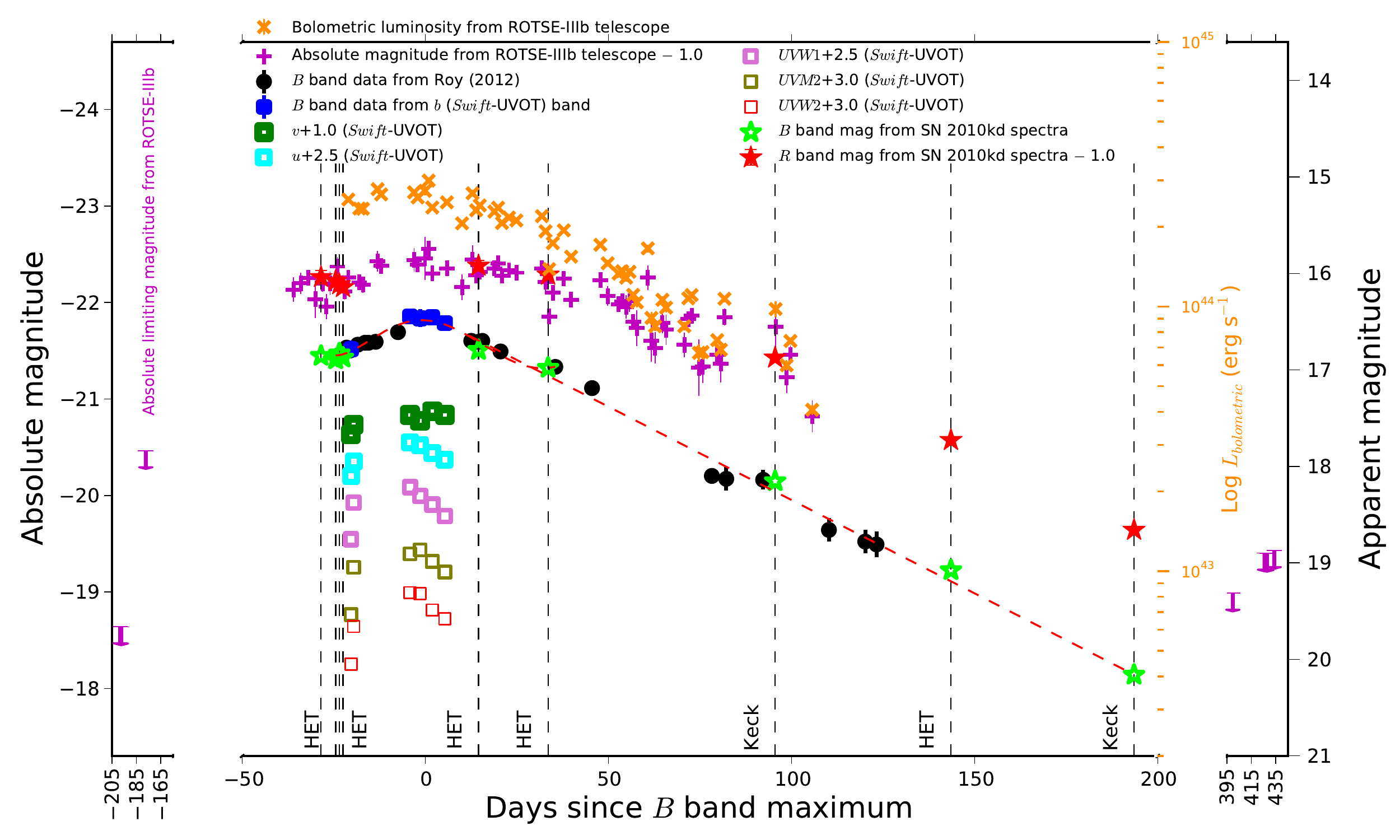}
\caption{Multiband light-curves of SN 2010kd, including the ROTSE-IIIb discovery photometric data in the 
clear filter (magenta plus signs) along with {\it Swift}-UVOT data (squares of different colors), are plotted. The Johnson \textit{B}-band photometry (in black) was taken from \cite{Roy2012}. The magenta arrows represent 
upper limits obtained by the ROTSE-IIIb at prediscovery and very late epochs.
The \textit{B} and \textit{R}-band magnitudes calculated from spectra of SN 2010kd using the {\tt sms} \citep{Inserra2018c} code are plotted with star symbols (lime and red, respectively). Around peak brightness, the $M_B$ light-curve is fitted with a third-order spline function to get the $M_{B, {\rm peak}}$ and the corresponding MJD$_{B,{\rm peak}}$. The postpeak (from peak to +194 days) data points are fitted using a straight line (in red dashed) to estimate the postpeak decay rate. Vertical dashed black lines represent epochs of spectral observations used in the present analysis. The bolometric light-curve calculated from ROTSE-IIIb discovery photometry is plotted in orange cross signs, showing a peak at $\sim (2.67 \pm 0.20) \times 10^{44}$ erg s$^{-1}$ (ROTSE-IIIb photometry data is published electronically).}
\label{fig:figBband}
\epsscale{1.}
\end{figure*}

\section{ROTSE-IIIb Discovery and Photometry} \label{sec:dis}
On 2010 November 14 (UT dates are used throughout this paper), SN 2010kd was discovered by the ROTSE-IIIb telescope as part of the ROTSE Supernova Verification Project (RSVP) at an apparent magnitude of $\sim 17.2$ mag \citep[]{Vinko2010}. SN 2010kd was discovered in a faint and metal-poor dwarf galaxy 
at J2000 coordinates $\alpha = 12^{\rm h}08^{\rm m}01\fs11$ and $\delta = +49\degr 13\arcmin 31\farcs1$. 

\begin{figure*}[ht!]
\includegraphics[angle=0,scale=0.92]{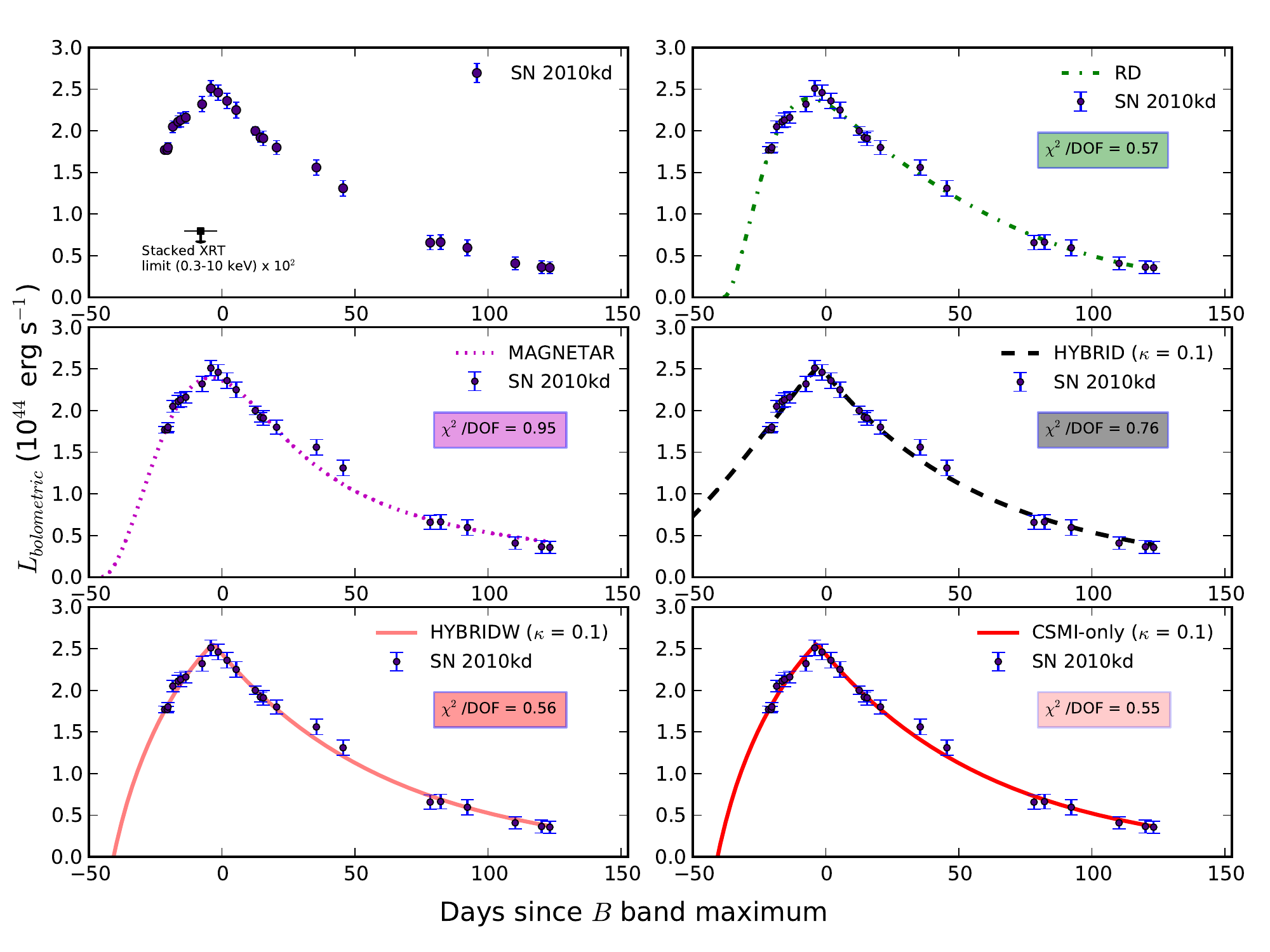}
\caption{In the upper-left panel, we plot the bolometric light-curve of SN 2010kd, whereas the semianalytic light-curve models (RD, MAG, HYBRID, HYBRIDW, and CSMI only) fitted to the bolometric light-curve of SN 2010kd using {\tt Minim} \citep{Chatzopoulos2013} are shown in five different panels. The photometric data points of SN 2010kd are shown in blue whereas the five modeled light-curves are color coded with the best-fit model (HYBRIDW/CSMI only) shown in red. The HYBRID, HYBRIDW, and CSMI only models are fitted with $\kappa = 0.1$ cm$^2$ g$^{-1}$. In the upper-left panel, the limiting XRT luminosity using stacked data (described in Appendix~\ref{XRTred}) is also shown with the black arrow.}
\label{fig:figMINIM}
\epsscale{6.}
\end{figure*}

The first optical spectrum of SN 2010kd was observed using the HET-9.2m on 2010 November 22, showing a narrow H$\alpha$ emission line from the host galaxy at redshift $z \approx 0.1$. There were no broad H or He features, but spectral modeling with the code {\tt SYNAPPS} \citep{Thomas2011} recognized an enrichment of carbon and oxygen lines \citep{Vinko2010,Vinko2012}. 
From follow-up photometry with ROTSE-IIIb, the peak absolute magnitude of SN 2010kd in the clear filter was constrained to be $\sim -21.4$ mag. The luminous peak absolute magnitude and the presence of O~II features at shorter wavelengths in the spectrum placed SN 2010kd in the category of SLSNe~I. Multiwavelength follow-up observations were triggered with the {\it Swift}-UVOT and XRT. {\it Swift}-UVOT detected SN 2010kd as a strong ultraviolet (UV) source. Even though no significant X-ray emission was detected, stacked observations gave an upper limit of $\sim 0.8 \times 10^{42}$ erg s$^{-1}$ (see the upper-left panel of Figure~\ref{fig:figMINIM}).
 
We constrained the explosion date to be MJD$_{\rm expl} \approx 55,483.09 \pm 0.90$ by extrapolating the premaximum light-curve down to the brightness level of the host galaxy ($m_u = +21.54 \pm 0.23$ mag; see Section~\ref{sec:host} for details) using a third-order spline function fit to the prepeak UV data. 
This calculated MJD$_{\rm expl}$ is in good agreement with that estimated by \cite{Chatzopoulos2013}. 
Throughout the present analysis, data of SN 2010kd were corrected for Galactic extinction, $E(B-V) = 0.02$ mag, using dust maps published by \cite{Schlegel1998}; however, the host-galaxy extinction, $E(B-V) = 0.15$ mag, was adopted from \cite{Schulze2018}. Throughout the paper, data have also been corrected for cosmological expansion; see Equation~\ref{eq:mag}.

\subsection{Optical Light-curve Analysis of SN 2010kd} \label{sec:Bband}
In Figure~\ref{fig:figBband}, {\it Swift}-UVOT photometry (\textit{UVW2}, \textit{UVM2}, \textit{UVW1}, \textit{u}, \textit{b}, and \textit{v} filters), 
ROTSE-IIIb photometry in the clear filter, and the \textit{B}-band data taken from \cite{Roy2012} are presented. Owing to the good temporal coverage of {\it Swift}-UVOT early-time data, the epoch of peak brightness could be tightly constrained in various bands. In Figure~\ref{fig:figBband}, we can see that the light-curve peak comes earlier in the UV bands and at later epochs in redder bands, as is typical of CCSNe \citep{Taddia2018}. We derived absolute magnitudes from the extinction-corrected apparent magnitudes using the formula
\begin{equation}\label{eq:mag}
M = m - 5\,\log(d_L\footnote{the luminosity distance}/10~{\rm pc}) + 2.5\,\log(1+z)
\end{equation}
\citep{Hogg2002,Lunnan2016}.
To precisely derive the \textit{B}-band peak brightness, {\it Swift}-UVOT \textit{b}-band data were converted to Johnson \textit{B} using the 
transformation equations given by \cite{Poole2008}. To constrain the late-time photometry (up to +194 
days), the \textit{B}- and \textit{R}-band magnitudes were calculated from the late-time spectra using the {\tt sms} \citep{Inserra2018c} code.
The derived values of synthetic spectral magnitudes and those estimated from the imaging are in good agreement (with a scatter $\lesssim 0.15$ mag). The \textit{B}-band maximum date (MJD$_{B,peak}$ = 55550.48) and the corresponding peak absolute magnitude ($M_{B,{\rm peak}} = -21.80 \pm 0.02$ mag) were calculated by fitting a third-order spline function to the $B$-band absolute magnitude ($M_B$) light-curve around maximum brightness ($-20$ to +20 days; red dashed line). The estimated value of the $M_{B,{\rm peak}}$ of the SN 2010kd is consistent with the mean peak absolute magnitudes of the sample of SLSNe~I published by \cite{Quimby2013}, \cite{Nicholl2016a}, and \cite{Inserra2018a}.

The postpeak light-curve (from peak to +194 days) can be fitted well with a single straight line (in mag; thus, exponential in flux) having a decay rate of $\sim 0.019$ mag day$^{-1}$, which is shallower than the
\isotope[56]{Ni} $\rightarrow$ \isotope[56]{Co} decay rate (0.11 mag day$^{-1}$) but steeper than that predicted for the \isotope[56]{Co} $\rightarrow$ \isotope[56]{Fe} decay rate (0.0098 mag day$^{-1}$). 
The $M_B$ light-curve decayed by 1 mag from the peak value in $\sim 56$ days, and the time taken to decay to the half-flux from the peak-flux value is $\sim 45$ days. Compared with the PTF sample of SLSNe~I \citep{DeCia2018}, the postpeak decay rate of SN 2010kd is shallower
than the average (0.04 mag day$^{-1}$) value of the early-time decline ($\lesssim +60$ days) but steeper 
than the average value ($\sim 0.01$ mag day$^{-1}$) of the late-time decline ($\gtrsim$ +60 days).

\begin{table*}[]
\scriptsize
 \begin{center}
  \begin{threeparttable}
    \caption{Best-fit Parameters for the RD Model.}
    \label{tab:table2}
    \addtolength{\tabcolsep}{16pt}
    \begin{tabular}{c c c c c c}

    \hline \hline

      $M_{\rm Ni}$ \tnote{a} & $t_d$ \tnote{b} & $A_{\gamma}$ \tnote{c} & $M_{\rm ej}$ \tnote{d}~(for $\kappa = 0.1$ cm$^2$ g$^{-1}$) & $\chi^2$/DOF \\

      ($M_\odot$) & (days) &  & ($M_\odot$) &  \\

      \hline

     20.3   $\pm$ 0.16 & 20.3  $\pm$  0.16 &  179.9   $\pm$   7.73 & 4.26 $\pm$ 0.13 & 0.57  \\
      \hline 
    \end{tabular}
    \begin{tablenotes}[para,flushleft]
        \item[a] $M_{\rm Ni}$: radioactive \isotope[56]{Ni} synthesized mass (in $M_\odot$).
        \item[b] $t_d$: effective diffusion timescale (in days).
        \item[c] $A_\gamma$: optical depth for the gamma-rays measured after the 10 days of explosion.
        \item[d] $M_{\rm ej}$: ejecta mass (in $M_\odot$).
    \end{tablenotes}
  \end{threeparttable}
 \end{center}
\end{table*}

\begin{table*}[]
\scriptsize
 \begin{center}
  \begin{threeparttable}
    \caption{Best-fit Parameters for the MAG Model.}
    \label{tab:table3}
    \addtolength{\tabcolsep}{-1pt}
    \begin{tabular}{c c c c c c c c c c}

    \hline \hline

     $R_0$ \tnote{a} & $E_p$ \tnote{b} & $t_d$ \tnote{c} & $t_p$ \tnote{d} & $v_{\rm exp}$ \tnote{e} & $M_{\rm ej}$ \tnote{c}~(for $\kappa = 0.1$ cm$^2$ s$^{-1}$)  & $P_i$ \tnote{f} & $B$ \tnote{g} & $\chi^2$/DOF \\

      ($10^{13}$ cm) & ($10^{51}$ erg) & (days) & (days) & ($10^3$ km s$^{-1}$) & ($M_\odot$) & (ms) & ($10^{14}$ G) &\\

      \hline

       1.0   $\pm$  1.25 & 3.37 $\pm$  0.06 & 35.0  $\pm$  0.88 & 46.85   $\pm$  2.08 & 11.2  $\pm$  0.98  & 5.71$\pm$ 0.285 & 2.44 $\pm$0.13 & 0.78$\pm$ 0.005 & 0.95 \\
      \hline 
    \end{tabular}
    \begin{tablenotes}[para,flushleft]
        \item[a] $R_0$: progenitor radius (in $10^{13}$ cm).
        \item[b] $E_p$: magnetar rotational energy (in $10^{51}$ erg).
        \item[c] discussed above.
        \item[d] $t_p$: magnetar spin-down timescale (in days).
        \item[e] $v_{\rm exp}$: SN expansion velocity (in $10^3$ km s$^{-1}$ ).
        \item[f] $P_i$: initial period of the magnetar (in ms).
        \item[g] $B$: magnetic field of the magnetar ($10^{14}$ G).
    \end{tablenotes}
  \end{threeparttable}
 \end{center}
\end{table*}

\begin{table*}[]
\scriptsize
 \begin{center}
  \begin{threeparttable}
    \caption{Best-fit Parameters for the HYBRID Model.}
    \label{tab:table4}
    \addtolength{\tabcolsep}{3.5pt}
    \begin{tabular}{c c c c c c c c c}  

    \hline \hline

     Opacity ($\kappa$) & $R_p$ \tnote{a} & $M_{\rm ej}$ \tnote{b} & $M_{\rm csm}$ \tnote{c} & $\dot{M}$ \tnote{d} & $M_{\rm Ni}$ \tnote{b} & $v_{\rm exp}$ \tnote{b} & $\chi^2$/DOF \\
      (cm$^2$ g$^{-1}$) & ($10^{13}$ cm) & ($M_\odot$) & ($M_\odot$) & ($M_\odot$ yr$^{-1}$) & ($M_\odot$) & ($10^3$ km s$^{-1}$) & \\
      \hline
       0.10 & 30.42  $\pm$ 1.57 & 19.51  $\pm$   0.22 & 29.56  $\pm$ 1.23 & 0.01  $\pm$  0.001  & 2.05   $\pm$  0.50 & 25.01  $\pm$  0.06 & 0.76 \\
      \hline 
    \end{tabular}
    \begin{tablenotes}[para,flushleft]
        \item[a] $R_p$: progenitor radius before the explosion (in $10^{13}$ cm).
        \item[b] discussed above.
        \item[c] $M_{\rm csm}$: CSM mass (in $M_\odot$).
        \item[d] $\dot{M}$: progenitor mass-loss rate (in $M_\odot$ yr$^{-1}$).
    \end{tablenotes}
  \end{threeparttable}
 \end{center}
\end{table*}

\begin{table*}[]
\scriptsize
  \begin{center}
    \begin{threeparttable}
    \caption{Best-fit Parameters for the HYBRIDW Model.}
    \label{tab:table5}
    \addtolength{\tabcolsep}{3.5pt}
    \begin{tabular}{c c c c c c c c c} 

    \hline \hline

     Opacity ($\kappa$) & $R_p$ \tnote{a} & $M_{\rm ej}$ \tnote{a} & $M_{\rm csm}$ \tnote{a} & $\dot{M}$ \tnote{a} & $M_{\rm Ni}$ \tnote{a} & $v_{\rm exp}$ \tnote{a} & $\chi^2$/DOF \\
      (cm$^2$ g$^{-1}$) & ($10^{13}$ cm) & ($M_\odot$) & ($M_\odot$) & ($M_\odot$ yr$^{-1}$) & ($M_\odot$) & ($10^3$ km s$^{-1}$) & \\

      \hline
       0.10 & 87.99 $\pm$    7.91 & 15.33   $\pm$  2.48 & 22.99 $\pm$   0.99 & 0.64 $\pm$  0.04 & 0.32   $\pm$ 0.26 & 24.99 $\pm$   0.27 & 0.56 \\
       
      \hline 
    \end{tabular}
    \begin{tablenotes}[para,flushleft]
        \item[a] Discussed above.
    \end{tablenotes}
  \end{threeparttable}
  \end{center}
\end{table*}

\begin{table*}[]
\scriptsize
  \begin{center}
    \begin{threeparttable}
    \caption{Best-fit Parameters for the CSMI only Model.}
    \label{tab:table6}
    \addtolength{\tabcolsep}{3.5pt}
    \begin{tabular}{c c c c c c c c c} 

    \hline \hline

     Opacity ($\kappa$) & $R_p$ \tnote{a} & $M_{\rm ej}$ \tnote{a} & $M_{\rm csm}$ \tnote{a} & $\dot{M}$ \tnote{a} & $M_{\rm Ni}$ \tnote{a} & $v_{\rm exp}$ \tnote{a} & $\chi^2$/DOF \\
      (cm$^2$ g$^{-1}$) & ($10^{13}$ cm) & ($M_\odot$) & ($M_\odot$) & ($M_\odot$ yr$^{-1}$) & ($M_\odot$) & ($10^3$ km s$^{-1}$) & \\

      \hline
       0.10 & 79.89 $\pm$    4.54 & 18.47   $\pm$  0.65 & 22.68 $\pm$   0.53 & 0.62 $\pm$  0.02 & 0.0   $\pm$ 0.0 & 24.75 $\pm$  0.11 & 0.55 \\
       
      \hline 
    \end{tabular}
    \begin{tablenotes}[para,flushleft]
        \item[a] Discussed above.
    \end{tablenotes}
  \end{threeparttable}
  \end{center}
\end{table*}

\subsection{Pseudobolometric light-curve Using ROTSE-IIIb} \label{sec:ROTSElum}
ROTSE-IIIb absolute magnitudes cover a temporal phase of $-35$ to +105 days, showing a clear filter peak absolute magnitude $\sim -21.4$ mag; see Figure~\ref{fig:figBband}. To obtain a pseudobolometric light-curve with ROTSE-IIIb unfiltered broadband photometry, we first perform a calibration using \textit{UBVRI} photometry from \cite{Roy2012}. Although there is no infrared (IR) photometry of SN 2010kd, a second calibration to \textit{UBVRIJHK} is performed by applying a correction from ROTSE-IIIb photometry to \textit{UBVRIJHK} for SN 2007gr \citep{Bianco2014}, another SN~Ic with extensive data in the UV through IR bands. For SN 2007gr, a linear interpolation is used to incorporate the flux between the \textit{I} and \textit{J}, \textit{J} and \textit{H}, and \textit{H} and \textit{K} filters. Although this is an approximation for the flux from SN 2010kd over IR wavelengths, a pseudobolometric light-curve is established from $-$22 to +105 days.

Owing to the fact that ROTSE-IIIb uses an open CCD with broad transmission covering 3000--10,000\,\AA, we first establish a calibration relation with \textit{UBVRI} photometry. Common epochs are derived by interpolating \textit{UBVRI} datasets to ROTSE-IIIb data. To obtain the flux, we calibrate ROTSE-IIIb magnitudes to \textit{R}-band magnitudes, correct for the extinction using values of $E(B-V) = 0.17$ mag, and use the Johnson-Cousins photometric system as described by \cite{Bessell1990}. Following the methodology of \cite{Dhungana2016}, a linear correlation is seen between the ratio of $L_{\rm ROTSE}$ and $L_{UBVRI}$ to the $(B-V)$ color. This linear relationship is then used to calibrate the $L_{\rm ROTSE}$ to $L_{UBVRI}$, establishing a pseudobolometric light-curve with peak luminosity $\sim (2.26 \pm 0.11) \times 10^{44}$ erg s$^{-1}$. The root-mean-square (RMS) of the residuals in fitting the $(B-V)$ dependence of $L_{\rm ROTSE}/L_{UBVRI}$ yields 8\% precision.

The same process is performed for SN 2007gr using publicly available $UBVr'i'$ data from the Fred L. Whipple Observatory (FLWO) on Mount Hopkins in Arizona \citep{Bianco2014}, similarly establishing a calibration with 7\% precision. To obtain the flux for this object, the Johnson system is used for the \textit{UBV} filters and the Sloan Digital Sky Survey (SDSS) photometric system \citep{Fukugita1996} is used for the $r'i'$ filters. Integrated luminosity is calculated by adopting a distance of 10.6 Mpc \citep{Chen2014}. Once ROTSE-IIIb is calibrated to \textit{UBVRI}, a second calibration from ROTSE-IIIb to \textit{UBVRIJHK} is performed by fitting the ratio of $L_{UBVRI}$ and $L_{UBVRIJHK}$ to the $(B-V)$ color and applying this to $L_{\rm ROTSE}$. The FLWO sample also contains \textit{JHK} data, where the Two Micron All Sky Survey system \citep{Cohen2003} is used to obtain flux for these filters. The RMS of the residuals for this calibration yields 6\% precision. In order to incorporate unobserved flux between the \textit{I} and \textit{J}, \textit{J} and \textit{H}, and \textit{H} and \textit{K} filters, a similar calibration is performed, this time fitting the ratio of $L_{UBVRIJHK}$ to the $L_{UBVRIJHK}$ including unobserved flux between filters versus the $(B-V)$ color. To calculate this additional flux at each epoch, a linear interpolation between the fluxes seen in each filter is used. Finally, the resulting fit is applied to $L_{\rm ROTSE}$ for SN 2010kd, increasing the peak luminosity to $\sim (2.67 \pm 0.20) \times 10^{44}$ erg s$^{-1}$, as seen in Figure~\ref{fig:figBband}.

\subsection{Bolometric light-curve and Model Fitting Using {\tt Minim}} \label{sec:MINIM}
Bolometric light-curve of the SN 2010kd was calculated using a Python-based code {\tt Superbol} \citep{Nicholl2018}. To get data near the peak brightness, we converted {\it Swift}-UVOT \textit{u}, \textit{b}, and \textit{v} magnitudes to Johnson \textit{U}, \textit{B}, and \textit{V} magnitudes, respectively, using the transformation equations given by \cite{Poole2008}. The \textit{R} and \textit{I} light-curves are interpolated using third-order polynomial fits to a common set of observed epochs with respect to observations in the \textit{B}-band. However, we assume a constant color to extrapolate late-time \textit{UVW2}, \textit{UVM2}, \textit{UVW1}, and \textit{U}-band data. We further extrapolate the blackbody (BB) spectral energy distribution (SED) by integrating the observed UV-optical flux to estimate the expected flux values in the near-infrared (NIR) region. The data were also corrected for Galactic as well as host extinction. The flux and wavelength of individual bands are shifted to the rest frame. The full bolometric light-curve covers phases from $-$20 to +123 days and shows a peak luminosity of $\sim (2.51 \pm 0.10) \times 10^{44}$ erg s$^{-1}$ (see upper-left panel of Figure~\ref{fig:figMINIM}). It is notable that the peak bolometric luminosity calculated using the method described in Section~\ref{sec:ROTSElum} from ROTSE-IIIb data and that derived with {\tt Superbol} for multiband data are in good agreement within the uncertainties.

We used the code {\tt Minim} \citep{Chatzopoulos2013} to fit various models to the bolometric light-curve of SN 2010kd. {\tt Minim} is a $\chi^2$-minimization code which utilizes Price algorithm \citep{Brachetti1997}, a controlled random search technique, briefly discussed by \cite{Chatzopoulos2013}. Semianalytic light-curve models --- RD, MAG, HYBRID (constant density CSMI + RD), HYBRIDW (wind-like CSMI + RD), and HYBRIDW with CSMI contribution only (CSMI only) models --- were fitted to the bolometric light-curve of SN 2010kd (see Figure~\ref{fig:figMINIM}).
The electron-scattering opacity ($\kappa$) may vary from $\sim 0.1$ to 0.2 cm$^2$ g$^{-1}$ for half- and fully-ionized materials, respectively \citep{Nicholl2015a}. We choose $\kappa = 0.1$ cm$^2$ g$^{-1}$, considering that in the case of H-poor SLSNe~I species like oxygen, carbon, and iron are roughly half ionized, as adopted by \cite{Inserra2013}. So, HYBRID, HYBRIDW, and CSMI only models are fitted with $\kappa = 0.1$ cm$^2$ g$^{-1}$.

All of the calculated parameters for the RD, MAG, HYBRID, HYBRIDW, and CSMI only models are, respectively, listed in Tables~\ref{tab:table2}$-$\ref{tab:table6}. Values of the $M_{\rm ej}$ are calculated using Equation~3 from \cite{Chatzopoulos2013}, both for the RD and the MAG models.

Statistically, all the models reproduced the bolometric light-curve of SN 2010kd within the error bars, as can be inferred from the respective values of $\chi^2$ per degree of freedom (DOF). But, the value of $M_{\rm Ni}$ given by the RD model is higher than the estimated $M_{\rm ej}$ value, which is unphysical; this indicates that the RD of \isotope[56]{Ni} cannot be considered as a primary power source of SN 2010kd. In comparison to other discussed models, the MAG model reproduced the bolometric light-curve of SN 2010kd with a higher value of $\chi^2$/DOF (0.95). However, the parameters of SN 2010kd estimated by the MAG model are in good agreement with the values suggested for PTF12dam \citep{Nicholl2013} and SN 2015bn \citep{Nicholl2016a}; see Table~\ref{tab:table3}.

The HYBRID and HYBRIDW models reproduced the bolometric light-curve of SN 2010kd with $\chi^2$/DOF = 0.76 and 0.56, respectively. In the lower-right panel of Figure~\ref{fig:figMINIM}, we also plot the HYBRIDW model with only a CSMI contribution by setting $M_{\rm Ni}$ = 0.0 $M_\odot$ (CSMI only model). The CSMI only model has also been able to reproduce the bolometric light-curve with the comparatively lower value of $\chi^2$/DOF = 0.55, which might indicate that \isotope[56]{Ni} heating does not play a significant role in the HYBRIDW (CSMI+RD) model.

With the lowest value of $\chi^2$/DOF (0.55), the CSMI only model constrains the $M_{\rm ej} \approx 18.47 \pm 0.65~M_\odot$, $M_{\rm csm} \approx 22.68 \pm 0.53~M_\odot$, and explosion ejecta velocity $\sim (24.75 \pm 0.11) \times 10^3$ km s$^{-1}$. The estimated $M_{\rm ej}$ value for SN 2010kd is higher in comparison to the median of the observed $M_{\rm ej}$ ($\sim 6\,M_\odot$) calculated using a sample of SLSNe~I \citep{Nicholl2015a} and a sample of SNe IIb, Ib, and Ic \citep[$\lesssim 4.4 \pm 1.3~M_\odot$;][]{Wheeler2015}. The parameters for SN 2010kd obtained using the best-fit CSMI only model are given in Table~\ref{tab:table6}. In summary, based on our model fitting, the CSMI or spin-down millisecond magnetar might be the possible powering mechanism for SN 2010kd.

Along with the best-fit bolometric light-curve, the limiting X-ray luminosity ($\lesssim 0.8 \times 10^{42}$ erg s$^{-1}$) calculated using the stacked images obtained from {\it Swift}-XRT is also plotted (upper-left panel of Figure~\ref{fig:figMINIM}, black arrow); data reduction is discussed in Appendix~\ref{XRTred}. SN 2010kd is not visible in X-rays even near peak optical brightness, as seen in other XRT-observed SLSNe~I except for PTF12dam \citep[which was detected;][]{Margutti2018}.

\begin{table*}[]
\scriptsize
  \begin{center}
    \caption{List of Well-studied SLSNe~I at Comparable Redshift Used for the Photometric Comparison with SN 2010kd.}
    \label{tab:table1}
    \addtolength{\tabcolsep}{-2pt}
    \begin{tabular}{c c c c c c c c c c c} 

    \hline \hline

     { } &  SLSN~I & R.A. ($\alpha$) & Decl. ($\delta$) & Redshift& $E(B-V)$& $E(B-V)$\footnote{For all the tabulated SLSNe~I, the host galaxy extinction values are taken from \cite{Schulze2018} except for Gaia16apd \citep{Kangas2017}.} & MJD$_{B,{\rm peak}}$ & $M_{B,{\rm peak}}$ & Source \\

      $ $ & & ($^h$:$^m$:$^s$) & ($^\circ:':''$) & ($z$) & (mag;Galactic)& (mag;Host) & & (mag) & & \\
      \hline

      1 & SN 2007bi & 13:19:20.19 &+\,08:55:44.3 & 0.128 & 0.02 & 0.04 &54159.47 & $-$21.41 $\pm$ 0.09\footnote{$M_{R, {\rm peak}}$ is taken in place of $M_{B, {\rm peak}}$.} &  \cite{Gal-Yam2009,Young2010}&\\

      2 & PTF10hgi/SN 2010md & 16:37:47.00 & +\,06:12:32.3 & 0.098& 0.07 & 0.01 & 55367.43	 & $-$20.66 $\pm$ 0.04 & \cite{Inserra2013,DeCia2018} &\\

     3 & SN 2011ke/PTF11dij/PS1-11xk & 13:50:57.77 & +\,26:16:42.8&0.143& 0.01 & 0.00 &55684.77 &$-$21.04 $\pm$ 0.15 & \cite{Inserra2013}&\\

      4 & SSS120810:231802-560926& 23:18:01.80  &$-$\,56:09:25.60 & 0.156 & 0.05 & 0.00 & 56158.30 & $-$21.29 $\pm$ 0.06 & \cite{Nicholl2014}&\\

    5 & PTF12dam & 14:24:46.20 &+\,46:13:48.3 & 0.107& 0.01 & 0.02 & 56093.70 & $-$21.68 $\pm$ 0.07 & \cite{Nicholl2013,Chen2015}&\\

    6 & SN 2015bn  &  11:33:41.57&+\,00:43:32.2 & 0.114 &0.02 &0.30 & 57103.38& $-$23.21 $\pm$ 0.10 & \cite{Nicholl2016a}&\\

    7 & Gaia16apd/SN 2016eay  & 12:02:51.71 & +\,44:15:27.4&0.101& 0.01 &0.01 &57549.59 & $-$21.86 $\pm$ 0.05 & \cite{Kangas2017,Nicholl2017}&\\

      \hline
    \end{tabular}

  \end{center}

\end{table*}

\section{Photometric Comparison of SN 2010kd with Other SLSNe~I} \label{sec:ligcomp}
For the present study, the following two broad criteria were adopted to build the sample of SLSNe~I  listed in Table~\ref{tab:table1}:

(1) The value of the redshift should be within $\pm\, 0.05$ relative to the redshift of SN 2010kd.

(2) The object should have photometric data in at least four bands (\textit{BVRI} or \textit{griz}).
 
For consistency, throughout this work the distances are calculated using the cosmological parameters $H_0 = 67.8 \pm 0.9$ km s$^{-1}$ Mpc$^{-1}$ , $\Omega_m$ = 0.308 $\pm$ 0.012, and $\Omega_\lambda$ = 0.692 \citep{PlanckCollaboration2016}. Except for Gaia16apd \citep{Kangas2017}, the host-galaxy extinction values for all the tabulated SLSNe~I are taken from \cite{Schulze2018}, which were estimated using Spectral Energy Distribution (SED) modeling. The $M_{B, {\rm peak}}$ and corresponding MJD$_{B,{\rm peak}}$ presented in Table~\ref{tab:table1} are calculated independently for all the tabulated SLSNe~I. SN 2007bi does not have \textit{B}-band magnitudes around the peak brightness, so MJD$_{R,{\rm peak}}$ is taken in place of MJD$_{B,{\rm peak}}$.

We can subdivide our sample of seven comparison SLSNe~I into fast-, slow-, and intermediate-decaying SLSNe~I based on postpeak decay rates. PTF10hgi, SN 2011ke \citep{Inserra2013}, and SSS120810:231802-560926 \citep{Nicholl2014} are designated as fast-decaying SLSNe~I, having postpeak decay rates close to that of \isotope[56]{Ni} $\rightarrow$ \isotope[56]{Co}. On the other hand, SN 2007bi \citep{Gal-Yam2009}, PTF12dam \citep{Nicholl2013}, and SN 2015bn \citep{Nicholl2016a} are slow-decaying SLSNe~I, having a postpeak decay rate close to that of \isotope[56]{Co} $\rightarrow$ \isotope[56]{Fe}. Photometrically, Gaia16apd resembles the an intermediate (between slow- and fast-decaying) SLSN, while its late-time spectroscopic features are similar to those of PTF12dam \citep{Kangas2017}. However, this division between fast and slow-decaying SLSNe~I is not broadly accepted for a larger sample of SLSNe~I \citep{Nicholl2017,DeCia2018,Lunnan2018}.

\subsection{\textit{B}-band Light-curve Comparison of SN 2010kd with Other SLSNe~I} \label{sec:ligcompMBBVRI}

\begin{figure}[ht!]
\includegraphics[angle=0,scale=0.70]{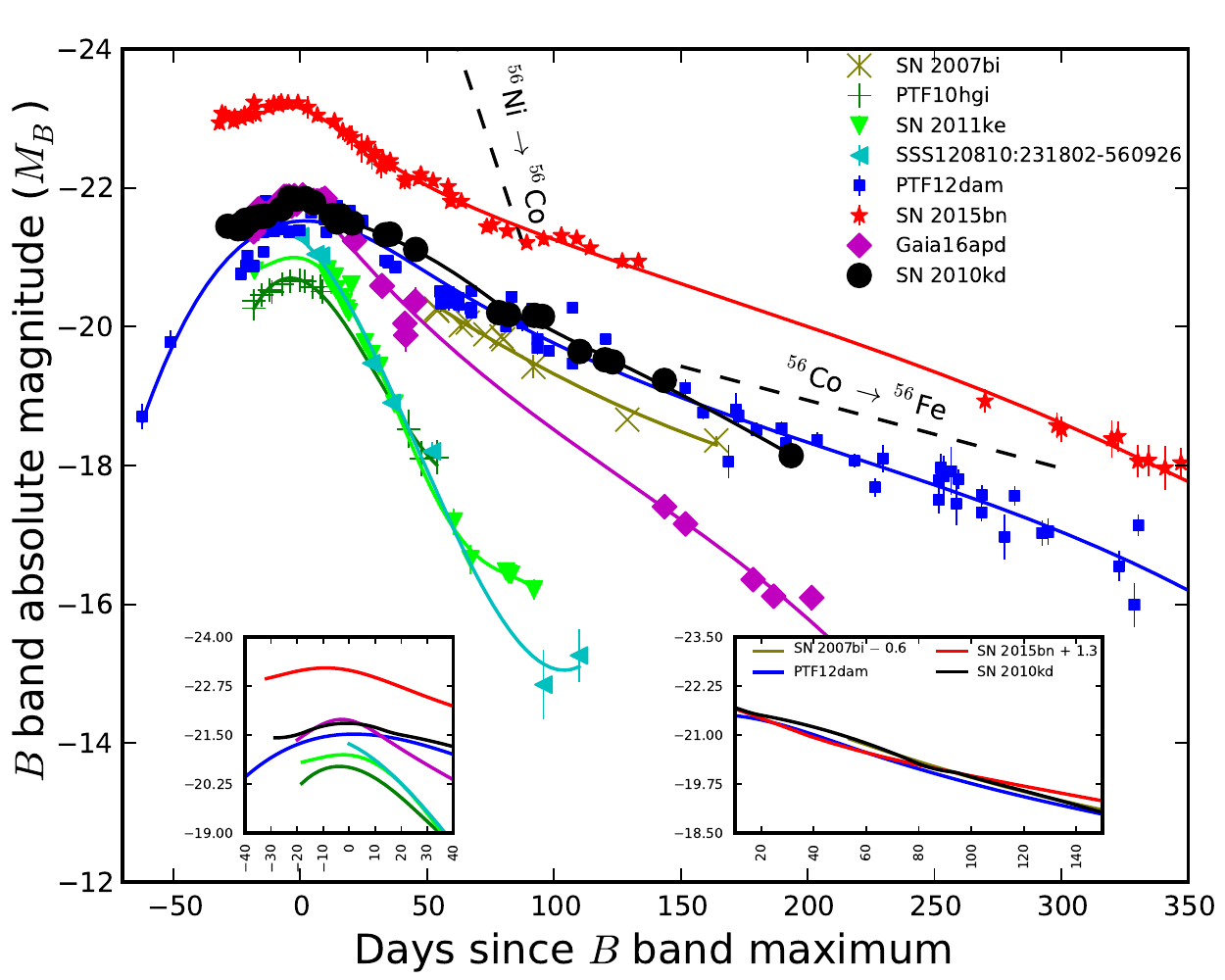}
\caption{The \textit{B}-band absolute magnitudes of SN 2010kd (in black) are compared with a subset of well-studied SLSNe~I at comparable redshift, as listed in Table~\ref{tab:table1}. We also compare these light-curves with the \isotope[56]{Ni} $\rightarrow$ \isotope[56]{Co} and \isotope[56]{Co} $\rightarrow$ \isotope[56]{Fe} theoretical decay curves, shown with black dotted lines. The \textit{B}-band light-curve of SN 2010kd nearly traces the decay pattern shown by SN 2007bi, PTF12dam, and SN 2015bn, representing slow-decaying SLSNe~I.}
\label{fig:figligcomp}
\epsscale{4.}
\end{figure}

In this section, we compare the $M_B$ light-curve of SN 2010kd with the light-curves of seven SLSNe~I given in Table~\ref{tab:table1} (see Figure~\ref{fig:figligcomp}). The light-curves of all the plotted SLSNe~I are corrected for the Galactic as well as host extinction. For those events having data in SDSS filters (SN 2011ke, PTF12dam, and SN 2015bn), the \textit{g}- and \textit{r}-band data were transformed to the \textit{B}-band using the transformation equations given by \cite{Jordi2006}. To trace the evolution, the $M_B$ light-curves of all the plotted SLSNe~I are fitted with a high-order spline function, shown with solid lines of different colors. SN 2010kd seems to have an $M_{B,{\rm peak}}$ value close to that seen in the case of PTF12dam and Gaia16apd. However, $M_{B,{\rm peak}}$ value of the SN 2010kd seems lower than those observed in case of SN 2015bn but larger than that observed for PTF10hgi, SN 2011ke, and SSS120810:231802-560926 (slow-decliners), as shown in the inset panel on the left side of Figure~\ref{fig:figligcomp}. This illustrates that SN 2010kd is one of the luminous SLSNe~I.

Fast-decaying SLSNe~I (PTF10hgi, SN 2011ke, and SSS120810:231802-560926) have decay rates of $\sim 0.06$ mag day$^{-1}$, slow-decaying SLSNe~I (SN 2007bi, PTF12dam, and SN 2015bn) show decay rates of $\sim 0.02$ mag day$^{-1}$, and Gaia16apd has a decay rate of $\sim 0.045$ mag day$^{-1}$. SN 2010kd has a decay rate of $\sim 0.019$ mag day$^{-1}$, similar to those estimated for slow-decaying SLSNe~I, especially matching with the decay pattern shown by PTF12dam, as shown in the inset panel on the right side of Figure~\ref{fig:figligcomp}. SN 2010kd seems to have a slow rising rate similar to SN 2015bn. All fast-decaying SLSNe~I in the present sample exhibit a decay rate slightly shallower than \isotope[56]{Ni} $\rightarrow$ \isotope[56]{Co}; however, all slow-decaying SLSNe~I of the sample exhibit a slightly steeper decay than
that predicted for \isotope[56]{Co} $\rightarrow$ \isotope[56]{Fe}. In summary, the \textit{B}-band absolute magnitude light-curve evolution of SN 2010kd is similar to that of slow-decaying SLSNe~I, and in particular closer to PTF12dam.

\subsection{Color Evolution Comparison of SN 2010kd with Other SLSNe~I} 
\label{sec:ligcompcolorB-V}
\begin{figure}[ht!]
\includegraphics[angle=0,scale=0.7]{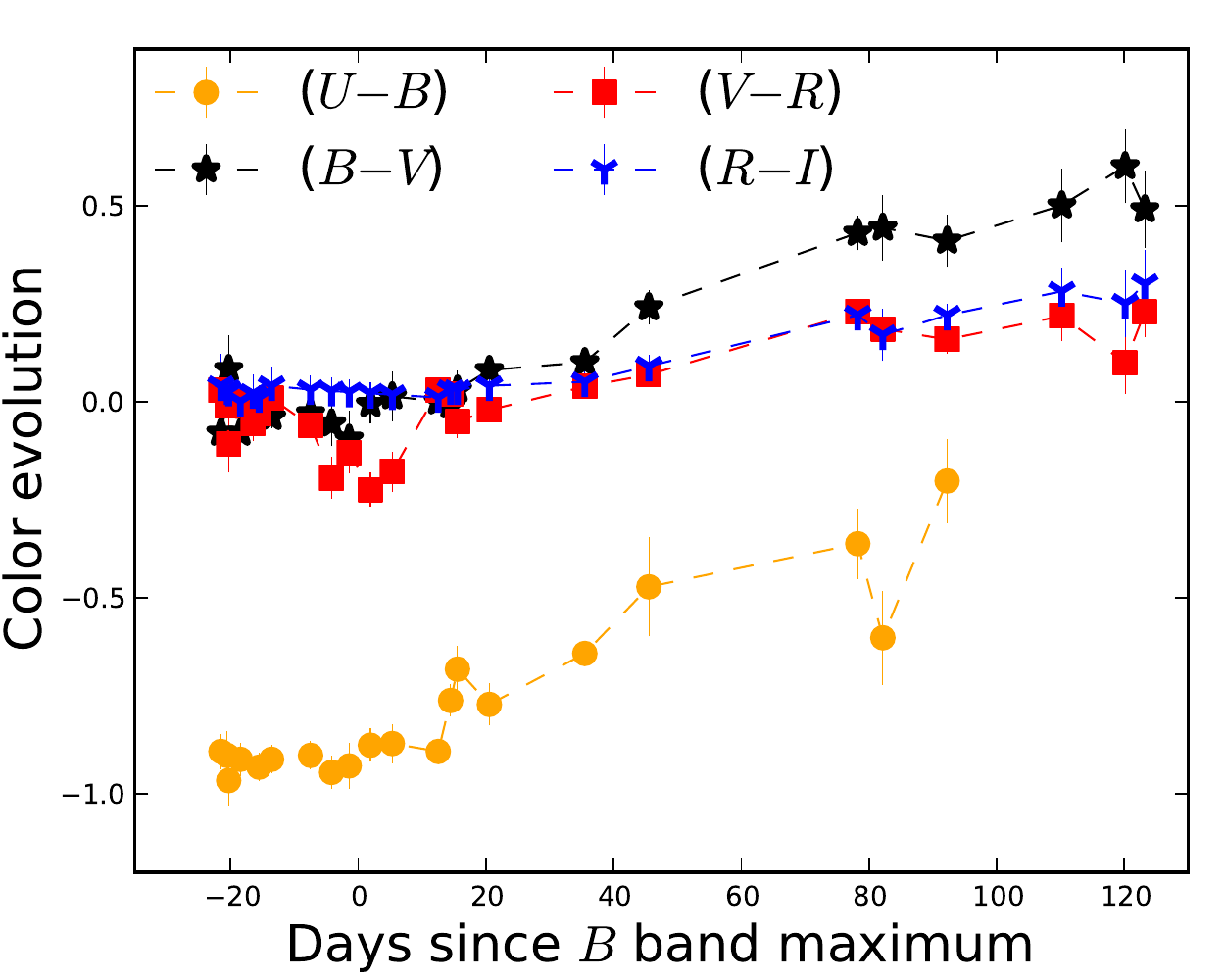}
\caption{The $(U-B)$, $(B-V)$, $(V-R)$, and $(R-I)$ color curves of SN 2010kd are plotted. The $(B-V)$, $(V-R)$, and $(R-I)$ color curves show nearly similar behavior and turn redder very slowly with respect to the $(U-B)$ color curve. The negative value of the $(U-B)$ color shows that SN 2010kd is brighter in the \textit{U}-band in comparison to the \textit{B}, \textit{V}, \textit{R}, and \textit{I} bands.}
\label{fig:figcolor}
\epsscale{3.}
\end{figure}

Optical through NIR colors of SLSNe~I are useful probes for understanding the temperature evolution during the photospheric-phase. The optical color evolution of SN 2010kd is plotted in Figure~\ref{fig:figcolor}. From $\sim -$20 to +15 days, all plotted color curves of SN 2010kd do not show significant change, indicating nearly constant temperatures around the time of peak brightness. After $\sim$+15 days, these color curves of SN 2010kd evolve redder, which can be associated with a decrease in the photospheric temperature as the SN dims during the postmaximum phases.

\begin{figure}[ht!]
\includegraphics[angle=0,scale=0.88]{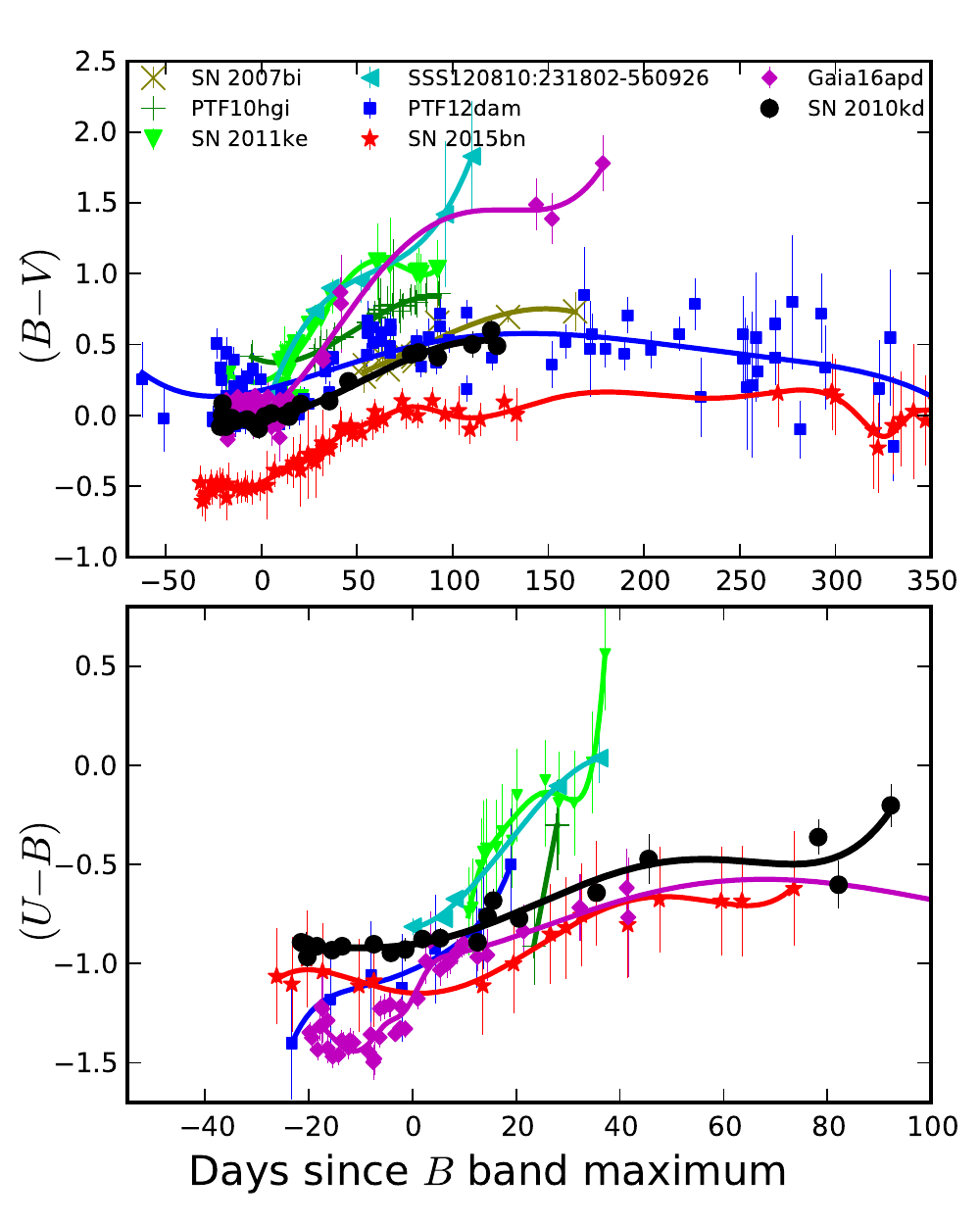}
\caption{Upper panel: the $(B-V)$ color curve of SN 2010kd is compared with the SLSNe~I listed in Table~\ref{tab:table1}. The $(B-V)$ color evolution of SN 2010kd becomes redder slowly in comparison to the fast-decaying SLSNe~I and appears to trace the path of slow-decliners. Lower panel: the $(U-B)$ color evolution of SN 2010kd is compared with the slow-decaying SLSNe~I, fast-decaying SLSNe~I, and one intermediate SLSN, Gaia16apd.}
\label{fig:figcompcolor}
\epsscale{5.}
\end{figure}

Figure~\ref{fig:figcompcolor} illustrates the evolution of the observed $(U-B)$ and $(B-V)$ color curves of SN 2010kd in comparison to SLSNe~I listed in Table~\ref{tab:table1}. Only the $(U-B)$ and $(B-V)$ colors are compared because $(V-R)$ and $(R-I)$ seem to evolve very similarly to $(B-V)$. Color curves of all the plotted SLSNe~I are corrected for the Galactic as well host-galaxy extinction and to trace the evolution, the color curves are fitted with a high-order spline function. For comparison, the SDSS $(u-g)$ and $(g-r)$ colors of PTF10hgi, SN 2011ke, PTF12dam and SN 2015bn were transformed to Johnson $(U-B)$ and $(B-V)$ colors, using the transformation equations and uncertainties given by \cite{Jordi2006}.

The $(B-V)$, $(V-R)$, and $(R-I)$ color curves appear to have nearly similar evolution, starting from $\sim 0$ (at $-20$ days) to $\sim +0.4$ mag (at +123 days), indicating slow cooling and expansion of the SN envelope. In the same time regime, the $(U-B)$ color  comparatively becomes redder ($\sim-$0.9 to $-$0.2 mag) sharply. This could be explained partly by cooling due to expansion and partly because of the enhancement of metallic (mostly Fe~II) features in the UV. Color curves of SN 2010kd show that it is very blue at early epochs (from +15 to $\sim +90$ days) and undergoes faster cooling in the \textit{U}-band compared with other optical bands.

The $(B-V)$ color evolution of fast-decaying (PTF10hgi, SN 2011ke, and SSS120810:231802-560926) SLSNe~I appears to become redder faster in comparison to slow-decliners (SN 2007bi, PTF12dam, and SN 2015bn); see the upper-panel of Figure~\ref{fig:figcompcolor}. It is also clear that for slow-decaying SLSNe~I, the $(B-V)$ color becomes redder slowly until $\sim 100$ days after peak brightness, and at later epochs, it becomes constant around $\sim 0.3$ mag (except SN 2015bn, which shows negative color values). In contrast, the $(B-V)$ colors of fast-decaying SLSNe~I become redder continuously up to $\sim 1.5$ mag (at $\sim 100$ days). The $(B-V)$ color evolution of Gaia16apd appears similar to the fast-decaying SLSNe~I. SN 2010kd seems to have a similar $(B-V)$ color curve to SN 2007bi and PTF12dam.

In the case of $(U-B)$ color comparison, SN 2007bi does not have data in \textit{U}-band, so the remaining six SLSNe~I are chosen for comparison (see Figure~\ref{fig:figcompcolor}, lower panel). Similar to $(B-V)$, the $(U-B)$ color evolution of the fast-decaying SLSNe~I turns redder sharply in comparison to the slow-decaying SLSNe~I. The $(U-B)$ color evolution of SN 2010kd seems similar to those seen in the case of Gaia16apd and SN 2015bn.

\begin{table*}[]
\scriptsize
 \begin{center}
  \begin{threeparttable}
    \caption{Log of the Spectroscopic Observations of SN 2010kd.}
    \label{tab:tablespec}
    \addtolength{\tabcolsep}{4pt}
    \begin{tabular}{c c c c c c c c c c} 

    \hline \hline

      Date & MJD &  Phase\footnote{Phase is given in days since \textit{B}-band maximum.} & Instrument & Wavelength  & Resolution & Exposure Time & Airmass &Telescope \\

      (UT) &  & (days) &  & (\AA) & (\AA) & (s) &   \\

      \hline

        2010 Nov 22 & 55,522.510 & $-$\,28 & LRS & 4200--9004 &  4.51 & 900 &1.24 &HET-9.2m \\
       
        2010 Nov 26 & 55,526.487 & $-$\,24 &  LRS & 4200--10,200  & 4.51 & 1200 &1.28& HET-9.2m \\
       
        2010 Nov 27 & 55,527.476 & $-$\,23 & LRS & 4200--10,200 & 4.51 & 900 &1.32 &HET-9.2m \\
       
        2010 Nov 28 & 55,528.513 &  $-$\,22 & LRS & 4200--10,200 & 4.51 & 900 &1.17& HET-9.2m \\
       
        2011 Jan 4 & 55,565.392 & +\,15 & LRS & 4200--10,200 & 4.51 & 1200 &1.24& HET-9.2m \\
       
       2011 Jan 23 & 55,584.353 & +\,34 &  LRS & 4200--10,200 & 4.51 & 1200 &1.20& HET-9.2m \\
       
       2011 Mar 26 & 55,646.284 & +\,96 &  LRIS & 3038--10,233 & 1.07 & 700 &1.30& Keck-10m \\
       
      2011 May 13 & 55,694.270 &  +\,144 & LRS & 4200--10,200 & 4.51 & 2200 &1.24& HET-9.2m \\
       
      2011 Jul 2 & 55,744.266 &  +\,194 & LRIS & 3250--10,100 & 1.07 & 970 &1.35& Keck-10m \\
       
      \hline 
 
    \end{tabular}
  \end{threeparttable}
 \end{center}
\end{table*}

\begin{figure*}[ht!]
\includegraphics[angle=0,scale=0.35]{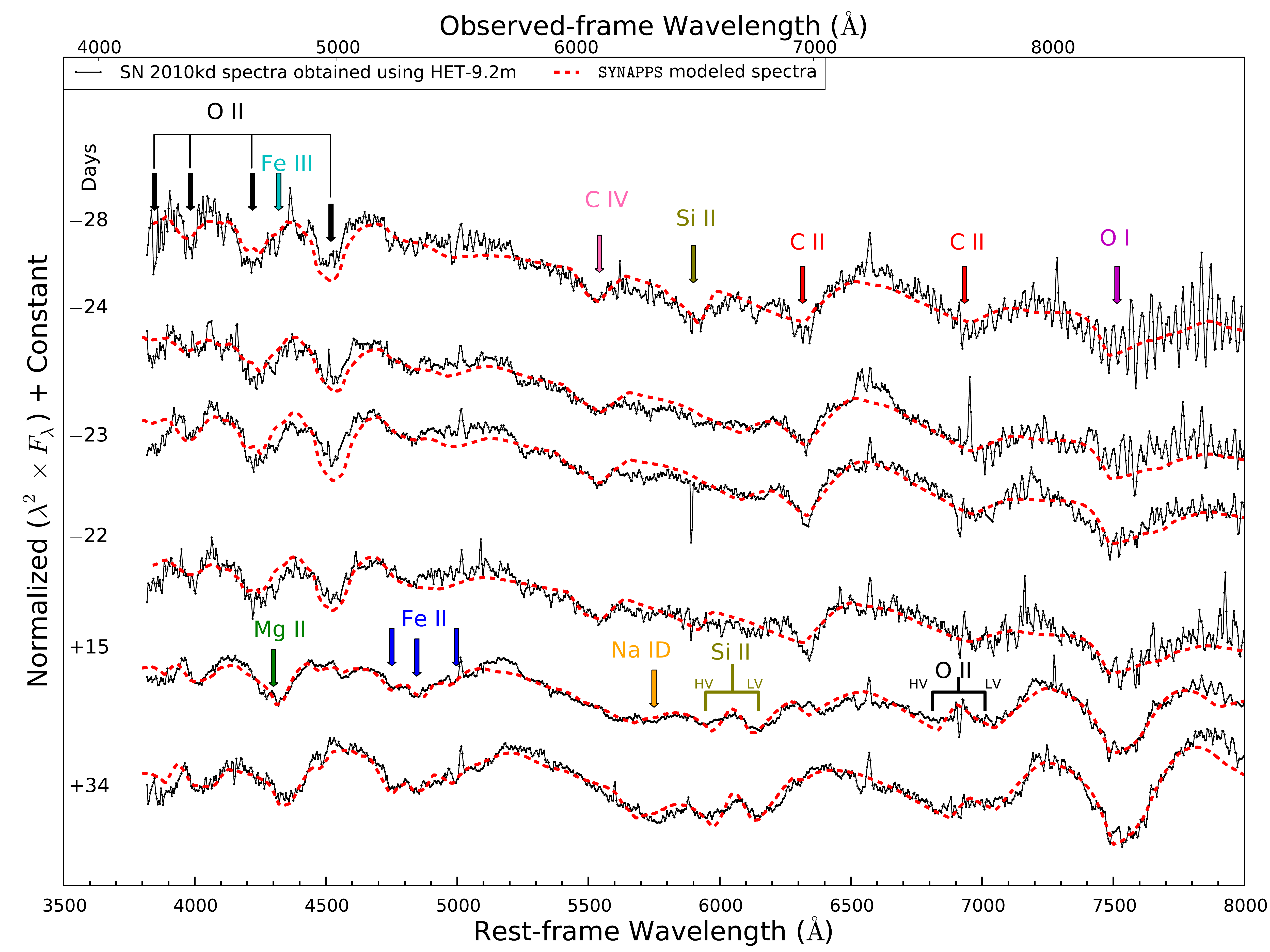}
\caption{The photospheric-phase spectral evolution of SN 2010kd (in black) along with the {\tt SYNAPPS} \citep{Thomas2011} spectral matching (in red) is plotted. Data have been corrected for the Galactic extinction and redshift. All spectral features are marked by vertical arrows at the observed-frame wavelengths of the elements, as given in their respective colors. The blue parts of the spectra are dense with O~II absorption lines. All six photospheric-phase spectra of SN 2010kd are well reproduced with the {\tt SYNAPPS} code.}
\label{fig:figphotphase}
\epsscale{7.}
\end{figure*}

\section{Spectroscopic Observations of SN 2010kd} \label{sec:spectroscopy}
As a part of the present analysis, spectra of SN 2010kd were acquired using the Marcario Low-Resolution Spectrograph \citep[LRS;][]{Hill1998} on the HET-9.2m and the Low-Resolution Imaging Spectrometer \citep[LRIS;][]{Oke1995} on the Keck-10m. A summary of the spectroscopic observations is given in Table~\ref{tab:tablespec}, with a coverage of $-$28 to +194 days. The HET-9.2m spectra were observed near the parallactic angle to minimize the effects of atmospheric dispersion. The Keck-10m spectra were obtained with an atmospheric dispersion corrector on LRIS, providing accurate relative spectrophotometry.

Data reduction was conducted with standard routines in $IRAF$\footnote{http://iraf.noao.edu/}, including bias and flat-field corrections, wavelength calibration, and flux calibration. At the HET-9.2~m, HgCd and ArNe spectral lamps are used for wavelength determination, while a spectrum of a suitable spectrophotometric standard star was taken on every night for performing the relative-flux calibration. The Keck-10m spectra were reduced in a similar manner. In this section, the symbol \AA\ is used for observed wavelengths and $\lambda$ for rest-frame wavelengths.

\begin{figure*}[ht!]
\includegraphics[angle=0,scale=0.35]{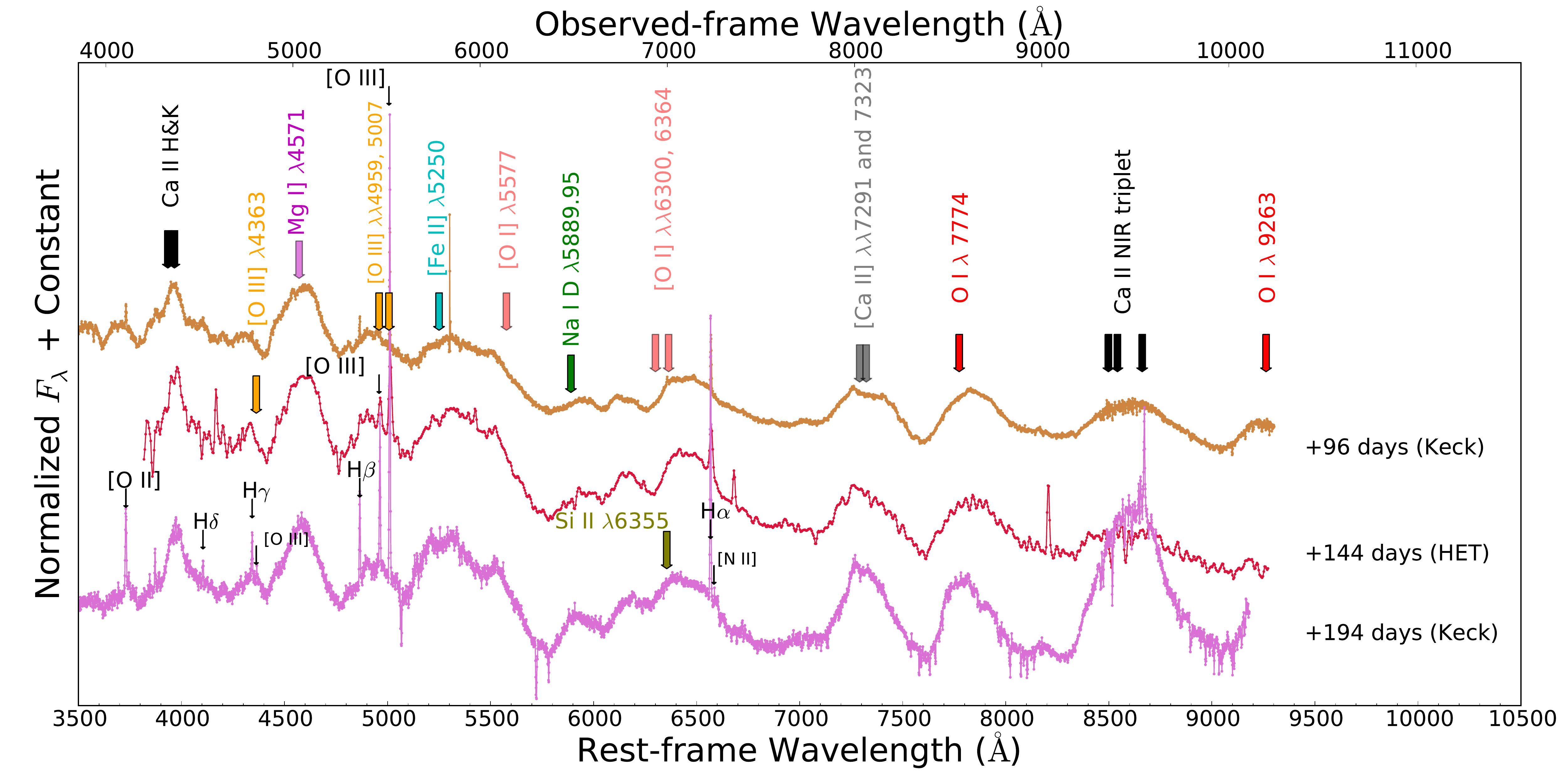}
\caption{The rest-frame spectral evolution of SN 2010kd in the nebular-phase (+96, +144, and +194 days) is plotted. Data have been dereddened and shifted to the rest-frame wavelengths. All spectral features are marked by vertical arrows at their rest-frame wavelengths, as represented by respective colors. In the spectrum at +194 days, the host-galaxy emission lines are marked with narrow black arrows. Nebular-phase spectra of SN 2010kd are dominated by O~I and Ca~II lines.}
\label{fig:fignebphase}
\epsscale{8.}
\end{figure*}

It is clear from Figures~\ref{fig:figphotphase} and \ref{fig:fignebphase} that the spectra of SN 2010kd could be considered in two parts. The first part consists of photospheric-phase spectra from $-$28 to +34 days; within this, four spectra belong to the hot photospheric-phase ($-$28, $-$24, $-$23, and $-$22 days) and two spectra belong to the cool photospheric-phase (+15 and +34 days). All spectra in the photospheric-phase were observed using the HET-9.2m. The second part covers three nebular-phase spectra observed at +96, +144, and +194 days. The spectrum at +144 days was observed using the HET-9.2m and the other two (+96 and +194 days) were observed with the Keck-10m.

\subsection{Photospheric-phase Spectra and {\tt SYNAPPS} Spectral Matching}\label{subsec:cont}
The photospheric-phase spectral evolution of SN 2010kd is plotted in Figure~\ref{fig:figphotphase}. Throughout the photospheric-phase, the spectra of SN 2010kd are dominated by the hot blue continuum, having a BB temperature range of $\sim 20,000-$ 10,000~K.
 
To explore possible identifications of the strong absorption lines, we modeled the first six spectra ($-$28, $-$24, $-$23, $-$22, +15, and +34 days) of SN 2010kd using the {\tt SYNAPPS} code \citep{Thomas2011}, which is a revised and improved version of SYNOW \citep{Jeffery1990}. We have modeled spectra in the photospheric-phase, because at this phase, the spectra have the most prominent features; moreover, {\tt SYNAPPS} only works in the photospheric-phase, not in the nebular-phase.

\begin{figure}[ht!]
\includegraphics[angle=0,scale=0.7]{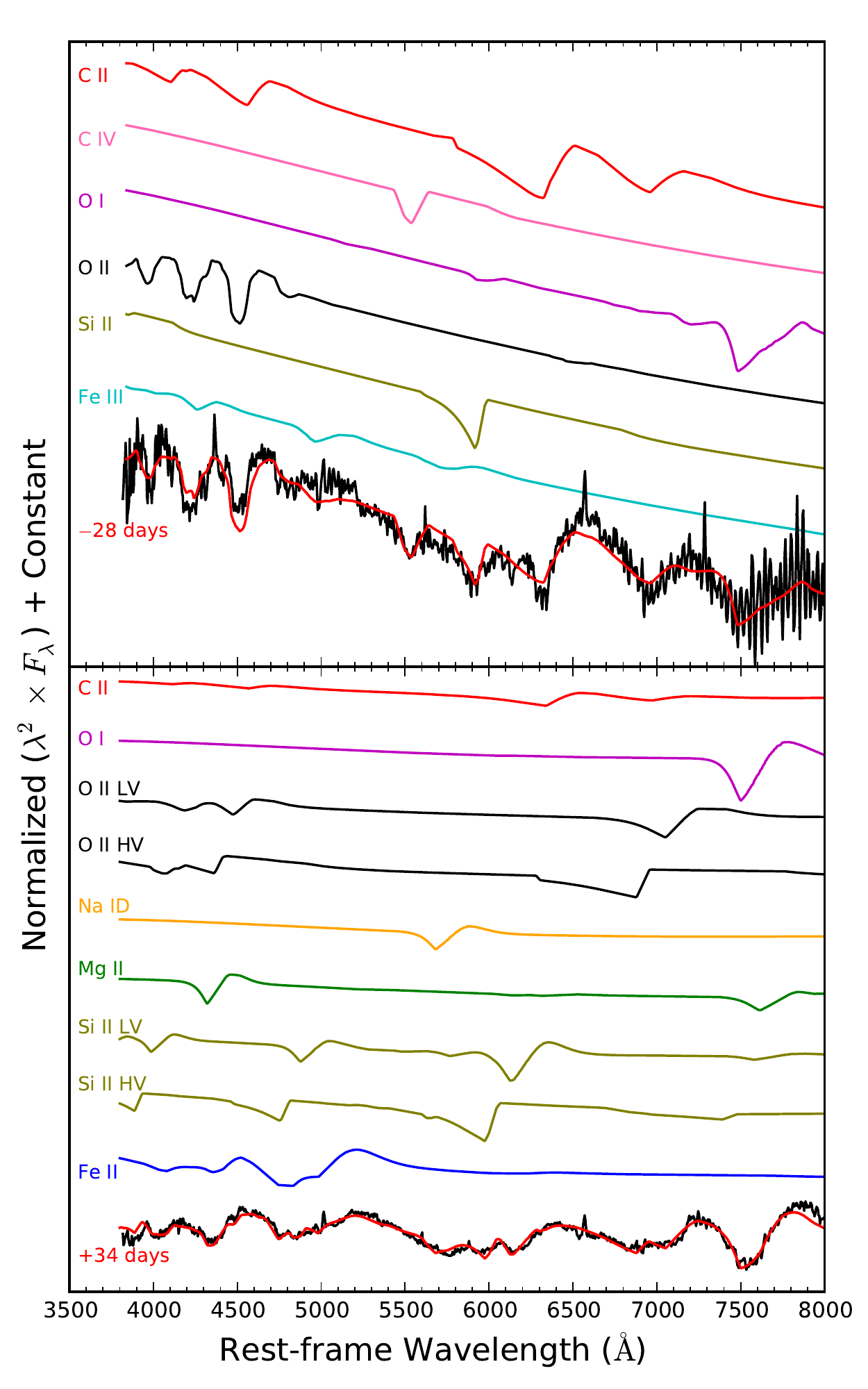}
\caption{Single-ion contributions to match the spectra of SN 2010kd at $-$28 and +34 days using the {\tt SYNAPPS} code are presented. In the spectrum at +34 days, the low- and high-velocity components of the O~II and Si~II lines are also found.}
\label{fig:figphotphaseindion}
\epsscale{7.}
\end{figure}

\begin{figure*}[ht!]
\includegraphics[angle=0,scale=0.45]{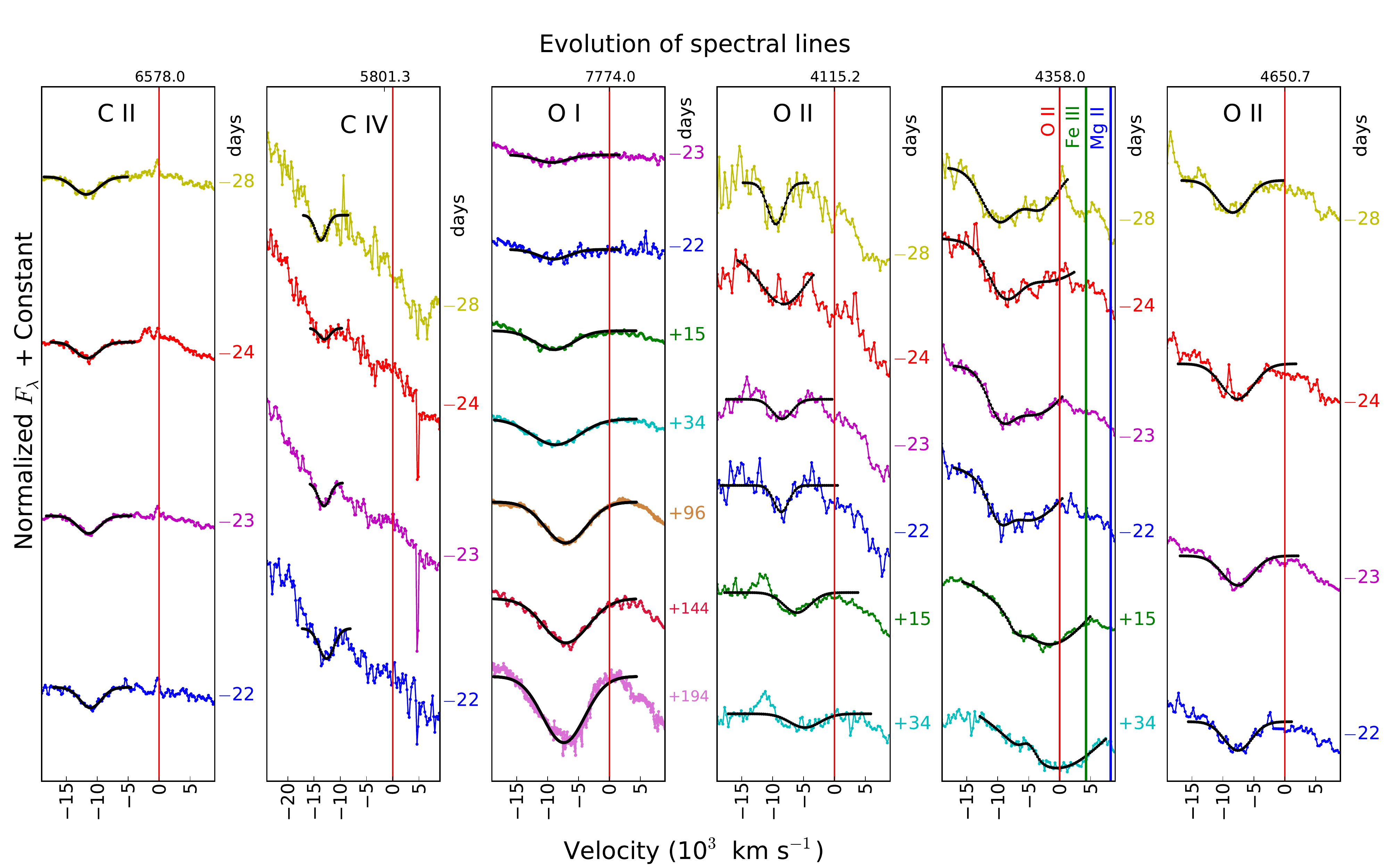}
\caption{Evolution of the C~II, C~IV, O~I, and O~II lines is plotted. Zero velocity is marked with a vertical line (in red), and the corresponding rest-frame wavelength is written on top. The phases of plotted spectra are $-$28, $-$24, $-$23, $-$22, +15, +34, +96, +144, and +194 days (from top to bottom). For every particular feature, the section of spectra is shown only for those phases in which it is visible. All lines are fitted with a single-Gaussian function (in black) except O~II $\lambda$4357.97, which seems to be blended with Mg~II and Fe~III lines and is fitted with a double Gaussian. All lines seem to have decreasing velocity as the phase increases.}
\label{fig:figelineevolution}
\epsscale{10.}
\end{figure*}

\begin{figure}[ht!]
\includegraphics[angle=0,scale=0.7]{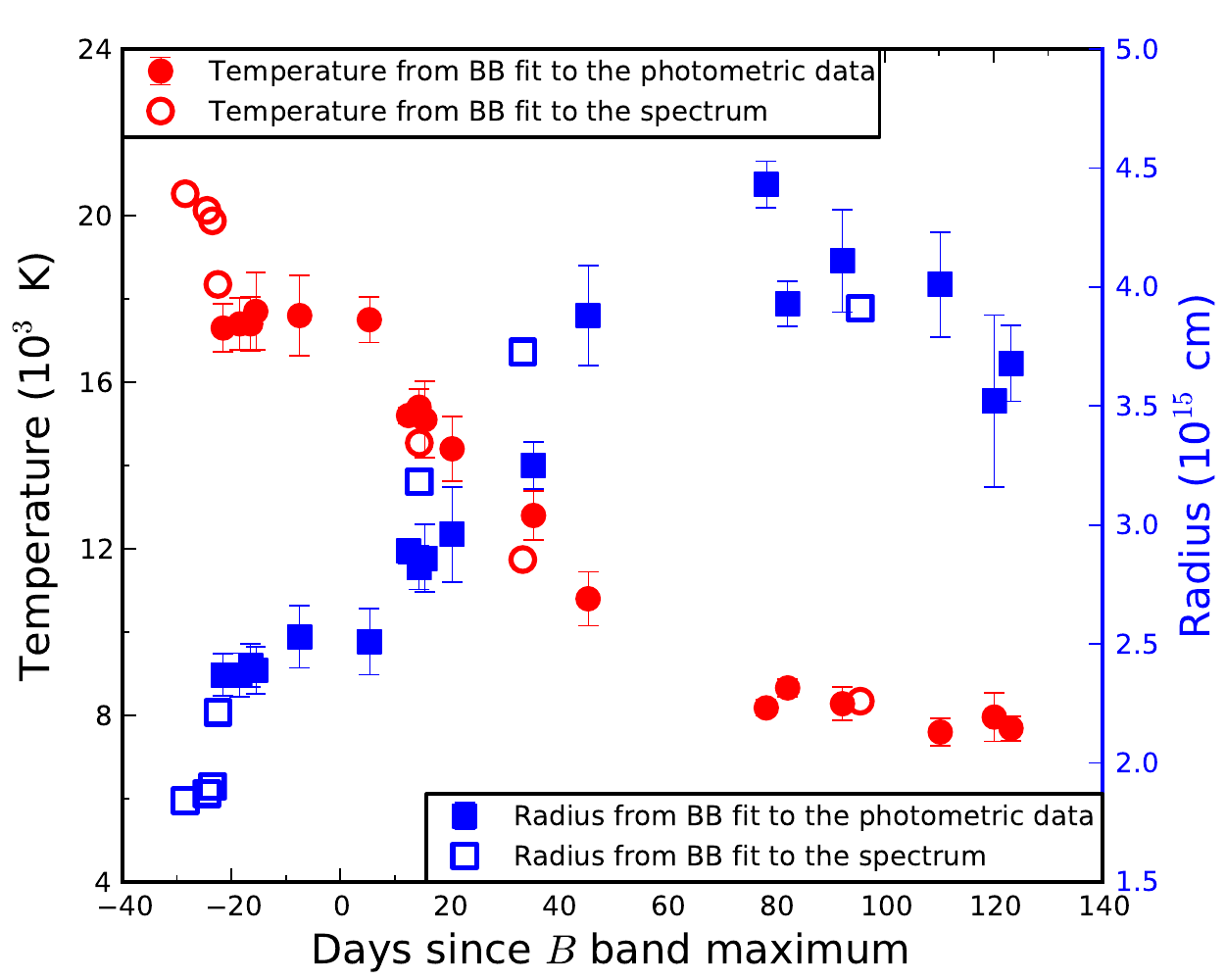}
\caption{Evolution of the BB temperature and radius of SN 2010kd as derived from the photometric data using {\tt Superbol} is plotted along with the temperature and the radius evolution calculated using the direct BB fit to the spectra. The temperature falls from $\sim 20,000$ to 8000~K from the hot photospheric to the nebular-phase, whereas the photospheric radius continuously increases from $-$28 to $\sim +50$ days and decreases slowly at later epochs.}
\label{fig:tempradifig}
\epsscale{4.1.}
\end{figure}

In Figure~\ref{fig:figphotphase}, we plot the observed spectra (in black) with the code-generated output synthetic spectra (in red). The ordinate plots $\lambda^2$ $\times$ flux density (${F_\lambda}$) to highlight the spectral matching at longer wavelengths. In photospheric-phase spectra at shorter wavelengths, many lines overlap, so their observed widths do not adequately represent the ejecta velocity. At longer wavelengths, on the other hand, line densities are less severe, and individual lines can be identified clearly.

Photospheric-phase spectra of SN 2010kd mainly exhibit C~II, C~IV, O~I, O~II, Na~I~D, Mg~II, Si~II, and Fe~II lines that are well reproduced by the code. No broad H or He lines are detected at any stage. Spectra in the hot photospheric-phase exhibit a strong C~II $\lambda$6578.05 line at  $\sim 6350$\,\AA\ which weakens in the cool photospheric-phase. In the hot photospheric-phase, the C~IV $\lambda$5801.31 line is also observed at $\sim 5530$\,\AA, with a consistent blueshifted absorption trough at a velocity around 14,000 km s$^{-1}$; this line disappears in the cool photospheric-phase.

It has been noticed that a large number of dense O~II lines is one of the strongest features in the visible-light spectra of SLSNe~I \citep[]{Quimby2011,Quimby2018}. The W-shaped O~II absorption lines are observed at $\sim 3980$, 4200, and 4500\,\AA, at a velocity of around 12,000 km s$^{-1}$; they are the dominant features of most SLSNe~I \citep[]{Quimby2011}. The O~II $\lambda$4357.97 line observed at $\sim 4200$\,\AA\ seems to have contributions from C~II and Fe~III in the hot photospheric-phase spectra. As temperature falls below $\sim 15,000$~K, the spectra fall into the cool photospheric-phase; O~II features weaken and are overtaken by heavier elements such as Mg~II and Fe~II \citep[]{Mazzali2016,Bose2018}.

Spectra at the redder end typically show a much weaker feature of O~I $\lambda$7774 at $\sim 7500$\,\AA, but it may have some contribution from Mg~II $\lambda$7877. Na~I~D absorption is not present in the spectra during the hot photospheric-phase but appear substantially in the cool photospheric-phase. The Si~II $\lambda$6355 line is strong in the spectrum at $-$28 days with a velocity around 21,000 km s$^{-1}$. Spectra at +15 and +34 days have low- and high-velocity components of Si~II and O~II, possibly indicating nuclear burning fronts in the outer layers of the explosion \citep{Hatano1999}. Lines detected at lower velocity ($\sim 11,500$ km s$^{-1}$) represent freshly synthesized material whereas higher velocity ($\sim 19,000$ km s$^{-1}$) components are likely to be primordial, as suggested by \cite{Hatano1999}.

In the hot photospheric-phase spectra, initially the Fe~II $\lambda5169$ lines are blended and dominated by the hot blue continuum, but as the spectrum evolves from the hot to the cool photospheric-phase, the Fe~II lines strengthen and are observed at $\sim 4500$--5200\,\AA. The contributions of individual ions to generate the spectra at $-$28 and +34 days are shown in Figure~\ref{fig:figphotphaseindion}.

\subsection{Evolution of Nebular-phase Spectra}\label{subsec:lines}
As SNe continue to expand and cool, their ejecta eventually become transparent, and they enter the nebular-phase. The nebular-phase begins much later in SLSNe~I ($\gtrsim +90$ days) compared to normal SNe~Ic, indicating that high densities are sustained for a longer time period in the massive ejecta of SLSNe~I \citep{Nicholl2015a}. The nebular spectra of SN 2010kd were taken at +96, +144, and +194 days (see Figure~\ref{fig:fignebphase}); line identification is based on that done by \cite{Inserra2017} and \cite{Jerkstrand2017}.
 
Ions and spectral features seen in the nebular-phase spectra of SN 2010kd are indicated with arrows of different colors at their respective rest-frame wavelengths. Dominant O~II and C~II absorption lines in the photospheric-phase spectra are now weakened or have disappeared in the nebular-phase; spectra are now dominated by O~I and Ca~II emission lines.

The region $\sim 5000$--5800\,\AA\ shows a broad and blended emission feature of [O~I] $\lambda$5577 and [Fe~II]. The region $\sim 5800$--6500\,\AA\ shows a trio of emission lines, possibly contributed by Na~I~D, Si~II, and the [O~I] $\lambda\lambda$6300, 6364 doublet. The O~I $\lambda$7774 line has been observed at a velocity around 7000 km s$^{-1}$; it is a recombination line that decays from the lower state of an O~I $\lambda$9263 transition. The nebular spectra of SN 2010kd exhibit a broad component of [O~III] lines which Section~\ref{sec:sec7} discusses in detail.

For SN 2010kd, the Mg~I] line begins to appear feebly after $\sim +34$ days and strongly in the later nebular-phases. The nebular-phase spectra of SN 2010kd show evolving Ca~II H and K, [Ca~II] $\lambda\lambda$7291, 7323, and the Ca~II $\lambda\lambda$8498, 8542, 8662 NIR triplet. The strength of the Ca~II, O~I, and Na~I~D absorption increases substantially in the nebular-phase spectra of SN 2010kd.

\subsection{Evolution of Spectral Lines of SN 2010kd}\label{sec:lineevolution}
Evolution of spectral features of SLSNe~I provides important clues to the interaction of CSM with the expanding ejecta and other important properties like geometrical distribution and dust formation in the ejecta. To highlight the evolution of individual lines, a section of spectra is plotted in the velocity domain corresponding to the rest-frame wavelength of various elements (see Figure~\ref{fig:figelineevolution}). The evolution of C~II $\lambda$6578, C~IV $\lambda$5801, O~I $\lambda$7774, and O~II $\lambda\lambda$4115.17, 4357.97, 4650.71 is presented. As the spectrum evolves, the minima of the absorption lines gradually move to lower velocities (i.e., toward redder wavelengths).

C~II and C~IV lines are present only in the hot photospheric-phase spectra; C~II seems prominent relative to C~IV and is well fitted with a single Gaussian. The C~II and C~IV line velocities estimated using the {\tt SYNAPPS} spectral matching (see Figure~\ref{fig:figvelomp}) and Gaussian fitting are in good agreement, indicating that these lines are free from blending, not seen in the case of many other lines.

As spectra evolve from the photospheric to the nebular-phase, the O~I $\lambda$7774 line evolves in FWHM intensity, depth, and width; it also becomes stronger and sharper, similar to those seen in the case of SN~Ic \citep{Pastorello2010}. The O~I $\lambda$7774 line is well fitted with a single Gaussian, but the estimated parameters are not reliable because of possible Mg~II $\lambda$7877 blending. The O~II lines seem to shift redward with phase, showing a decreasing velocity; however, velocity estimates using single-Gaussian fitting are not reliable when considering blended features. It is best to use {\tt SYNAPPS} spectral matching for blended features.

\subsection{Temperature and Radius Evolution of SN 2010kd}\label{sec:tempradifig}
In this section, we discuss the evolution of the BB temperature (in red) and radius (in blue) of SN 2010kd derived from the photometric and spectroscopic data (see Figure~\ref{fig:tempradifig}). We calculate the photospheric temperature and radius evolution by modeling the SED at possible epochs by fitting a BB function using the code {\tt Superbol} \citep{Nicholl2018}. The temperature and radius values estimated using the BB fit individually to the spectra are also shown. Overall, the temperature and radius evolution calculated from two different methods are in good agreement.

At premaximum phases (from $\sim -28$ to $-15$ days), the temperature of SN 2010kd decays sharply from $\sim 20,000$ to 17,000 K, whereas around the time of peak brightness (from $\sim -15$ to +15 days) the temperature seems nearly constant. After peak brightness, from $\sim +15$ to +80 days, the temperature decays from $\sim 15,000$ to 8000~K, and at later phases (after $\sim +80$ days), it decays shallower ($\sim 5$~K per day) relative to the decay rate at early phases.

From $\sim -28$ to +50 days, the radius of SN 2010kd sharply increases (from $\sim (1.8$ to $4.0) \times 10^{15}$ cm), whereas it decreases slowly ($\sim 0.03 \times 10^{15}$ cm per day) after $\sim +80$ days. The photospheric temperature of SN 2010kd during the earliest phases ($> 12,000$~K) seems to be somewhat hotter than that of PTF12dam and SN 2015bn ($\lesssim 12,000$~K; \citealt{Nicholl2016a}). The BB radius estimated for SN 2010kd is smaller in comparison to SN 2015bn (from $\sim (8$ to $12) \times 10^{15}$ cm) and decays comparatively shallower after +50 days \citep{Nicholl2016a}.

\section{Spectral Comparison of SN 2010kd with Other H-deficient SNe}\label{sec:comwithother}
SLSNe~I spectra near peak brightness exhibit differences in several features among the two populations i.e., slow- and fast-decaying 
\citep{Nicholl2016a,Quimby2018}. Hence, in the next two sections, we compare the spectra of SN 2010kd with the template spectra of slow- and fast-decaying SLSNe~I. To study the spectral diversity among SNe~Ic and SLSNe~I, we compared the spectra of SN 2010kd with a subset of well-studied broad-lined SNe~Ic.

\subsection{Comparison of SN 2010kd with Slow-decaying SLSNe~I}\label{sec:comwithothersl}
\begin{figure}[ht!]
\includegraphics[angle=0,scale=1.2]{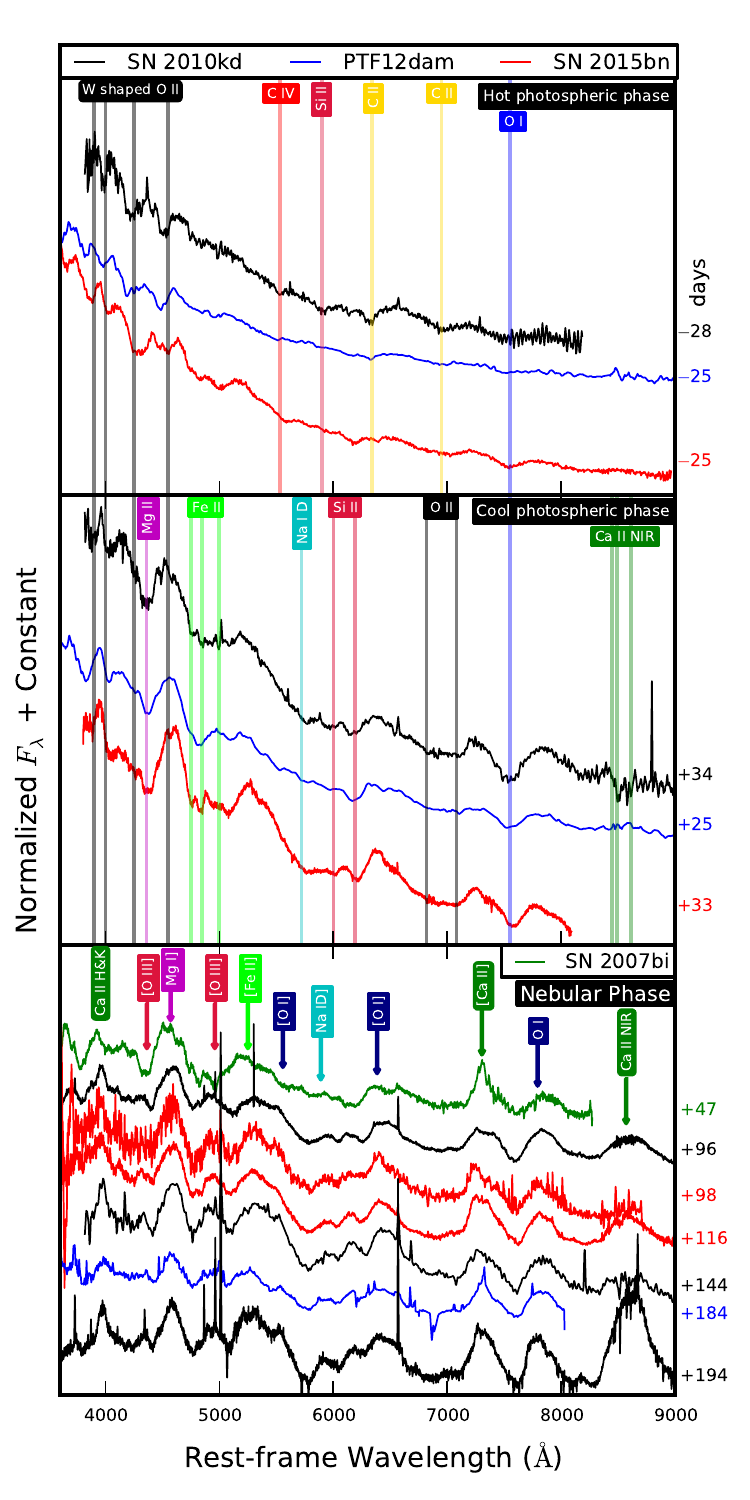}
\caption{Spectra of SN 2010kd and three slow-decaying SLSNe~I at possible similar phases are compared. Spectra of SN 2007bi are plotted in green, SN 2010kd in black, PTF12dam in blue, and SN 2015bn in red. The spectral evolution of SN 2010kd appears to be an excellent match to those of SN 2007bi, PTF12dam, and SN 2015bn in the photospheric as well as nebular-phases.}
\label{fig:figcompslow}
\epsscale{10.}
\end{figure}

In this section, observed spectra of SN 2010kd are compared with three slow-decaying SLSNe~I: SN 2007bi \citep[][in green]{Gal-Yam2009}, PTF12dam \citep[][in blue]{Nicholl2013}, and SN 2015bn \citep[][in red]{Nicholl2016a}; see Figure~\ref{fig:figcompslow}. All of the plotted spectra have been shifted to their respective rest-frame wavelengths. For a significant comparison, spectra of PTF12dam and SN 2015bn are chosen at epochs close to those of the spectra of SN 2010kd. As shown in Figure~\ref{fig:figcompslow}, the spectral evolution of SN 2010kd appears to be an excellent match to those of SN 2007bi, PTF12dam, and SN 2015bn; however, there are also some petty differences.

In the two upper panels of Figure~\ref{fig:figcompslow}, the photospheric spectra of SN 2010kd at $-$28 and +34 days are compared with PTF12dam and SN 2015bn. All of the colored vertical lines are plotted at the observed wavelengths of various elements, associated with their respective colors. In the photospheric-phase, the continuum temperature of SN 2010kd looks hotter than that of PTF12dam and SN 2015bn, as discussed in Section~\ref{sec:tempradifig}. The spectra of SN 2010kd and SN 2015bn are consistent for a longer time around maximum light than PTF12dam. This may be because of their shallower velocity decay rates in comparison to PTF12dam (see Figure~\ref{fig:figvelomp}).

Both SN 2010kd and PTF12dam have a C~II line around the same velocity \citep{Quimby2018}, whereas SN 2015bn only has a meager signature of this line \citep{Nicholl2016a}. SN 2015bn shows C~IV $\lambda$5801.31 weaker than SN 2010kd, but \cite{Nicholl2016a} claimed it as an unidentified line or a Si~II line of very high velocity. In early-time spectra, PTF12dam also exhibits a weak trough of this line at $\sim 5750$\,\AA, but \cite{Quimby2018} claimed this feature as an unidentified line, or He $\lambda$5876. SN 2010kd, PTF12dam, and SN 2015bn seem to have have similar O~I $\lambda$7774 line evolution, whereas the lack of O~I could be a distinguishing difference between the slow- and fast-decaying SLSNe~I \citep{Quimby2018}. The hot photospheric-phase spectra of SN 2010kd and PTF12dam have stronger O~II lines \citep{Quimby2018}, whereas in SN 2015bn these lines are comparatively weaker even at $-$25 days \citep{Nicholl2016a}. However, in the case of SN 2010kd and SN 2015bn, the O~II $\lambda$4357.97 line is blended with Fe~III $\lambda$4420, which is not significant in the case of PTF12dam \citep{Nicholl2016a}. The centers of absorption troughs show that the velocity of the O~II lines in SN 2015bn appear to be lower compared to SN 2010kd and PTF12dam.

In the lower panel of Figure~\ref{fig:figcompslow}, three nebular spectra of SN 2010kd are compared with the nebular spectra of SN 2007bi, PTF12dam, and SN 2015bn. The nebular-phase in SLSNe~I starts after $\sim +90$ days, but we also included the +47 day spectrum of SN 2007bi because it seems to have many similar lines (except [O~III] $\lambda$5007) as in the +96 day spectrum of SN 2010kd.

For SN 2010kd, the blending of [O~I] with [Fe~II] lines seems to increase with phase, as can also be seen in spectra of SN 2007bi, PTF12dam, and SN 2015bn. SN 2010kd and SN 2015bn have stronger [O~III] $\lambda\lambda$4959, 5007 broad emission lines in comparison to PTF12dam, whereas [O~III] $\lambda$4363 is absent in the early-phase (at +47 days) spectrum of SN 2007bi.

In summary, high-cadence, early-time spectra exhibit a small degree of diversity among these objects, but spectroscopic evolution of SN 2010kd overall confirms the similarity of this object to SN 2007bi, PTF12dam, and SN 2015bn.

\subsection{Comparison of SN 2010kd with Fast-decaying SLSNe~I}\label{sec:compwithfast}
\begin{figure}[ht!]
\includegraphics[angle=0,scale=1.2]{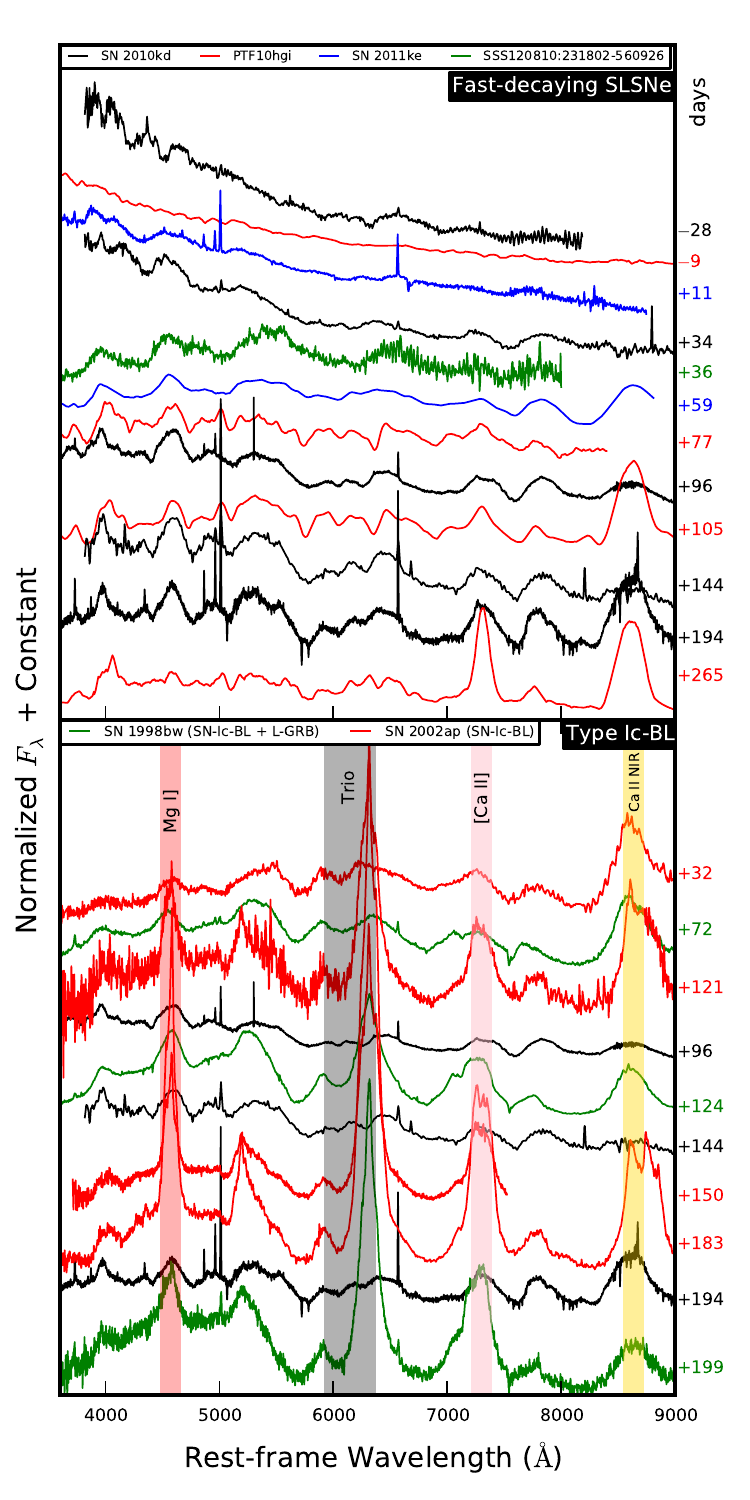}
\caption{Upper: spectral comparison of SN 2010kd with three well-studied fast-decaying SLSNe~I: PTF10hgi, SN 2011ke, and SSS120810:231802-560926. This comparison shows that fast-decaying SLSNe~I spectroscopically evolve faster in comparison to SN 2010kd. Lower: comparison of late-phase spectra of SN 2010kd with two well-studied SNe~Ic-BL: SN 1998bw and SN 2002ap. SN 2010kd and SNe~Ic-BL seem to have many similar lines but with different intensities.}
\label{fig:figcomp}
\epsscale{11.}
\end{figure}

In this section, we compare the same sequence of spectra of SN 2010kd with three well-studied fast-decaying SLSNe~I: PTF10hgi \citep[][in red]{Inserra2013}, SN 2011ke \citep[][in blue]{Inserra2013}, and SSS120810:231802-560926 \citep[][in green]{Nicholl2014}. For a significant comparison in the nebular-phase, we cover a range from +59 to +265 days. As can be seen from the upper panel of Figure~\ref{fig:figcomp}, SN 2010kd maintained a strong blue continuum for a much longer time relative to the fast-decaying SLSNe~I, which exhibit a redder continuum even at $\sim +36$ days.

The fast-decaying SLSNe~I seem to have smoother spectra in comparison to SN 2010kd, indicating broader velocity distributions which blend and smooth out individual lines \citep{Quimby2018}. 
SN 2010kd appears to have sharper absorption lines and higher emission-line luminosities in the nebular-phase with respect to the fast-declining SLSNe~I.

Despite the above-mentioned differences, both classes of SLSNe~I exhibit nearly the same lines and in the same order of strength, mainly calcium, Mg~I], and [Fe~II]. These lines become prominent at early phases (from $\sim +36$ days) in the fast-decaying SLSNe~I, but at later phases ($\sim +96$ days) in objects like SN 2010kd owing to their slower photometric evolution. Spectra of photometrically fast-decaying SLSNe~I seem to evolve faster in comparison to SN 2010kd. The result is that the spectra of fast-decaying SLSNe~I at $\sim +100$ days seem to differ from spectra of slow-decaying SLSN at $\sim +100$ days and are better matched to the slow-decliners at $\sim +200$ days, as also suggested by \cite{Quimby2018}. Slower spectral evolution of slow-declining SN 2010kd might indicate higher ejecta mass and more slowly decreasing velocities in comparison to the fast-declining SLSNe~I.


\subsection{Comparison of SN 2010kd with Broad-lined Type Ic SNe}\label{sec:compwithIc}
In this section, we compare the nebular-phase spectra of SN 2010kd (in black) with spectra of two well-studied Type Ic-BL SNe: SN 1998bw \citep[][in green]{Sollerman2000} and SN 2002ap \citep[][in red]{Vinko2004}; see the lower panel of Figure~\ref{fig:figcomp}. SN 1998bw is a broad-lined event associated with a long/soft GRB (SN~Ic-BL + LGRB), whereas SN 2002ap is a broad-lined SN~Ic with no associated GRB (SN~Ic-BL). SN 2002ap seems to evolve much faster than SN 1998bw and SN 2010kd. The spectrum of SN 2002ap at +32 days appears similar to the spectrum of SN 1998bw at +72 days and that of SN 2010kd at +194 days. The nebular-phase spectra of SN 2010kd seem to evolve slower in comparison to SN 1998bw and SN 2002ap, as studied in the case of other slow-decaying SLSNe~I \citep{Nicholl2016b,Jerkstrand2017}.
 
SN 1998bw and SN 2002ap exhibit similar major spectral features to SN 2010kd (except in the region $\sim 5700$--6500~\AA), but different behavior in terms of the shape and density of the line features. At phases $\sim +90$ to +200 days, in the range $\sim 5700$--6500~\AA, the most important line for comparison is the [O~I] doublet, generally the strongest feature in nebular spectra of SNe~Ic and not severely contaminated by other lines. In this range, SN 2010kd shows the trio of Si~II, Na~I~D, and the [O~I] doublet (shaded with gray color), which can be seen in nebular-phase spectra of most SLSNe~I (see Figure~\ref{fig:figcompslow}). As suggested by \cite{Nicholl2016a}, we can take this trio as a real spectroscopic difference between spectra of SLSNe~I and SNe~Ic. SN 2010kd reveals weak [O~I] $\lambda\lambda$6300, 6364 in all three nebular spectra, but this line could become very prominent at late phases (after $\sim 300$ days) as shown by \cite{Jerkstrand2017} for SN 2015bn. The broad [O~III] $\lambda$4363 and $\lambda\lambda$4959, 5007 emission lines are strong in SN 2010kd, whereas SN 1998bw and SN 2002ap do not have any evidence of these lines.

The other significant difference is in the calcium and magnesium spectral lines. SN 1998bw and SN 2002ap show strong emission lines of Mg~I], Ca~II H and K, [Ca~II], and the Ca~II NIR triplet compared to SN 2010kd (shaded with pink and gold colors). SN 2002ap has the strongest Mg~I] line in comparison to SN 1998bw and SN 2010kd. Spectra of SN 2002ap and SN 1998bw show a clear blending of the [O~II] $\lambda$7235 line with [Ca~II] up to late phases, whereas this blend is not clear in SN 2010kd.

All three SNe have [Fe~II] $\lambda$5250, but this line is again strongest in SN 2002ap. [Fe~II] is blended with [O~I] $\lambda$5577 in the spectra of SN 2010kd, which is not significant in plotted SNe~Ic-BL. The O~I $\lambda$7774 line is most prominent in SN 2010kd in comparison to SN 1998bw and SN 2002ap.

In summary, we can say that SN 2010kd evolves comparatively slower and seems to have many lines similar to those of SNe~Ic-BL but with different intensities. This indicates towards a clear diversity in the nebular-phase spectra of SLSNe~I and SNe~Ic-BL, an important key to understanding the diverse underlying physical mechanisms in different type of SNe~Ic.

\begin{figure*}[ht!]
\includegraphics[angle=0,scale=0.6]{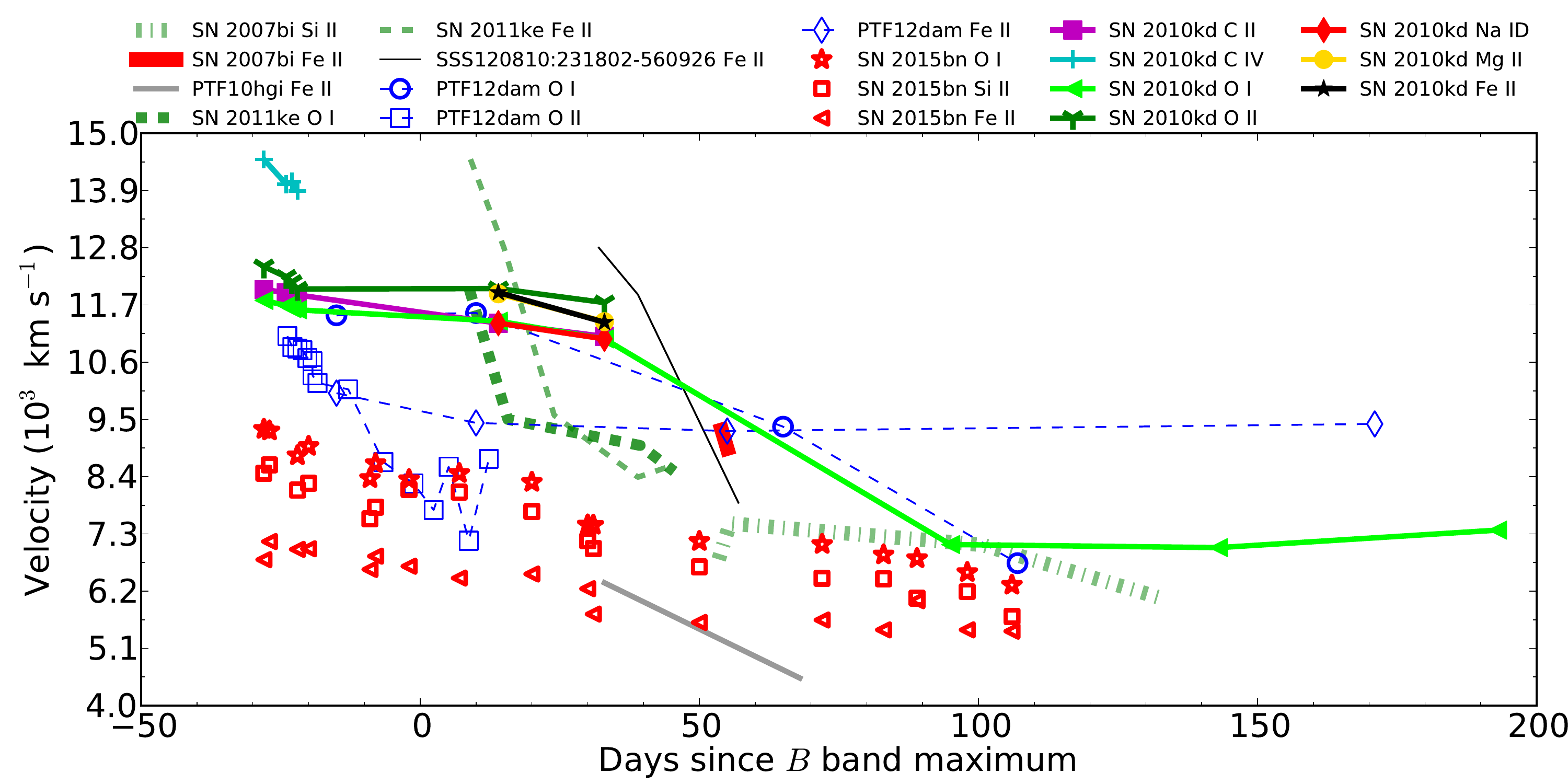}
\caption{The line velocities of SN 2010kd are compared with a set of well-studied SLSNe~I at comparable redshift. Measurements were made by {\tt SYNAPPS} spectral matching; owing to blending of different elements, the single-Gaussian-fitting method is not a good way to calculate velocities. In comparison to SN 2010kd, fast-decaying SLSNe as well as PTF12dam appear to have velocities with steeper decay rates, whereas SN 2015bn seems to have lower velocity values with similar decay rates. Data Sources: \cite{Young2010,Inserra2013,Nicholl2016a,Liu2017,Quimby2018}.}
\label{fig:figvelomp}
\epsscale{12.}
\end{figure*}

\section{Line-velocities Comparison of SN 2010kd with Other SLSNe~I}\label{sec:velocomp}
In this section, we determined line velocities of SN 2010kd using {\tt SYNAPPS} spectral matching and compared them with a set of well-studied SLSNe~I at comparable redshift; see Figure~\ref{fig:figvelomp}. Line velocities of SN 2007bi \citep{Young2010,Liu2017}, PTF10hgi \citep{Liu2017}, SN 2011ke \citep{Inserra2013,Liu2017}, SSS120810:231802-560926 \citep{Liu2017}, PTF12dam \citep{Nicholl2016a,Quimby2018}, and SN 2015bn \citep{Nicholl2016a} are plotted for comparison.

For SN 2010kd, the velocity evolution of the C~II, C~IV, O~I, O~II, Na~I~D, Mg~II, and Fe~II lines are plotted. In the photospheric-phase (from $-$28 to +34 days), all lines show nearly constant velocity curves having a range of $\sim 12,500-$11,000 km s$^{-1}$, except C~IV, which has the highest velocity around 14,000 km s$^{-1}$. In the photospheric-phase, the velocity of the O~I $\lambda7774$ line remains almost constant, which is commonly observed in other SLSNe~I as well \citep{Nicholl2015a,Nicholl2016a}; see Figure~\ref{fig:figvelomp}. The O~I line velocities at nebular-phases are calculated using single-Gaussian fitting, because we cannot use {\tt SYNAPPS} at such late phases. The constant velocity evolution possibly indicates stratification of line-forming shells within a homologous expansion. The C~II, C~IV, Na~I~D, Mg~II, and Fe~II lines also show a shallow decline in velocity with time ($\sim 200$ km s$^{-1}$ every 10 days).
In the case of SN 2010kd, the relation between the estimated velocity of the Fe~II $\lambda5169$ line at $\sim +10$ days ($\lesssim 12,000$ km s$^{-1}$) and its decay rate between $\sim +10$ and $\sim +30$ days ($\sim 30$ km s$^{-1}$ day$^{-1}$) is in good agreement with those observed in the case of other slow-decaying SLSNe~I \citep{Inserra2018b}.

The O~II line velocity of SN 2010kd is nearly constant ($\sim 12,000$ km s$^{-1}$) in comparison to PTF12dam, which has a declining O~II line velocity (from $\sim 11,000$ to 7000 km s$^{-1}$) with a rate of $\sim 1200$ km s$^{-1}$ per 10 days. The O~I line velocity of SN 2010kd nearly traces the path estimated for PTF12dam. 

The evolution of the Fe~II line velocity of SN 2010kd is almost flatter in comparison to the fast-decaying SLSNe~I (PTF10hgi, SN 2011ke, and SSS120810:231802-560926), which indicates that photometrically fast-decaying SLSNe~I have comparatively faster-declining Fe~II line velocities \citep{Inserra2018b}.

In summary, SN 2010kd has flatter velocity curves in comparison to SN 2007bi, PTF12dam, and other fast-decaying SLSNe~I, but similar velocity evolution to that seen in the case of SN 2015bn. The flat velocity curves of SN 2010kd might indicate possible signature of a central engine accelerating the inner ejecta as suggested by \cite{Nicholl2016a} for SN 2015bn.

\begin{figure*}[ht!]
\includegraphics[angle=0,scale=0.75]{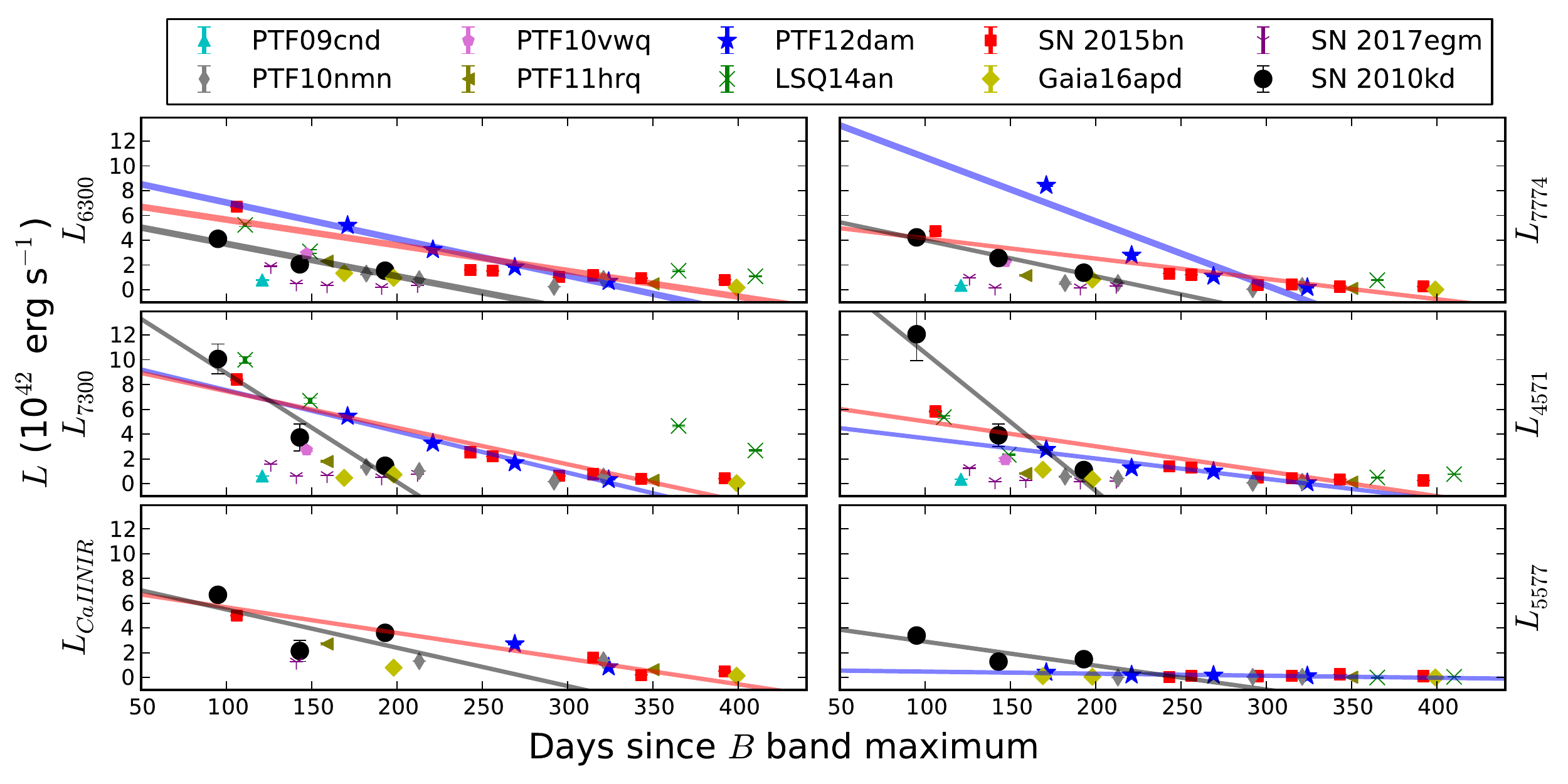}
\caption{Integrated emission-line luminosities of SN 2010kd compared with a set of well-studied SLSNe~I published in the literature. Line luminosities of SN 2010kd (in black), PTF12dam (in blue), and SN 2015bn (in red) are fitted with straight lines (as shown with respective colors) to make a rough estimate of line luminosities and line-luminosity decay rates. Most of the line luminosities of SN 2010kd seem to have higher values at +96 days as well as higher luminosity decay rates than other SLSNe~I. By comparing the [O~I] $\lambda\lambda$6300, 6364 line-luminosity of SN 2010kd with that of other SLSNe~I, we estimated an upper limit of O~I ejected mass of $\sim 10~ M_\odot$. Data sources: \cite{Nicholl2013,Nicholl2016a,Nicholl2016b,Nicholl2017,Nicholl2019,Chen2015,Inserra2017,Jerkstrand2017,Kangas2017,Quimby2018}.}
\label{fig:figlumcomp}
\epsscale{8.}
\end{figure*}

\section{Line Luminosities in the Nebular-phase}\label{sec:linelum}
In this section, we present integrated luminosities of various emission lines in nebular spectra of SN 2010kd: [O~I] $\lambda$5577, [O~I] $\lambda\lambda$6300, 6364, O~I $\lambda$7774, Mg~I] $\lambda$4571, [Ca~II] $\lambda\lambda$7291, 7323, and the Ca~II NIR triplet (see Figure~\ref{fig:figlumcomp}, in black). First, the continuum was subtracted by a linear fit, and the values of luminosities are estimated by direct numerical integration only for the lines that are isolated and do not have highly blended features, except [O~I] $\lambda$5577. For the [O~I] $\lambda$5577 line, we fitted a double Gaussian, because it is blended with [Fe~II] $\lambda$5250, the same method adopted by \cite{Nicholl2019}. We also attempted subtracting the contribution of the host-galaxy flux from the nebular-phase spectra of SN 2010kd using the host spectrum published by \cite{Leloudas2015} as described by \cite{Nicholl2019}; however, the host-galaxy continuum flux of SN 2010kd is found to be negligible.

\begin{figure*}[ht!]
\includegraphics[angle=0,scale=0.75]{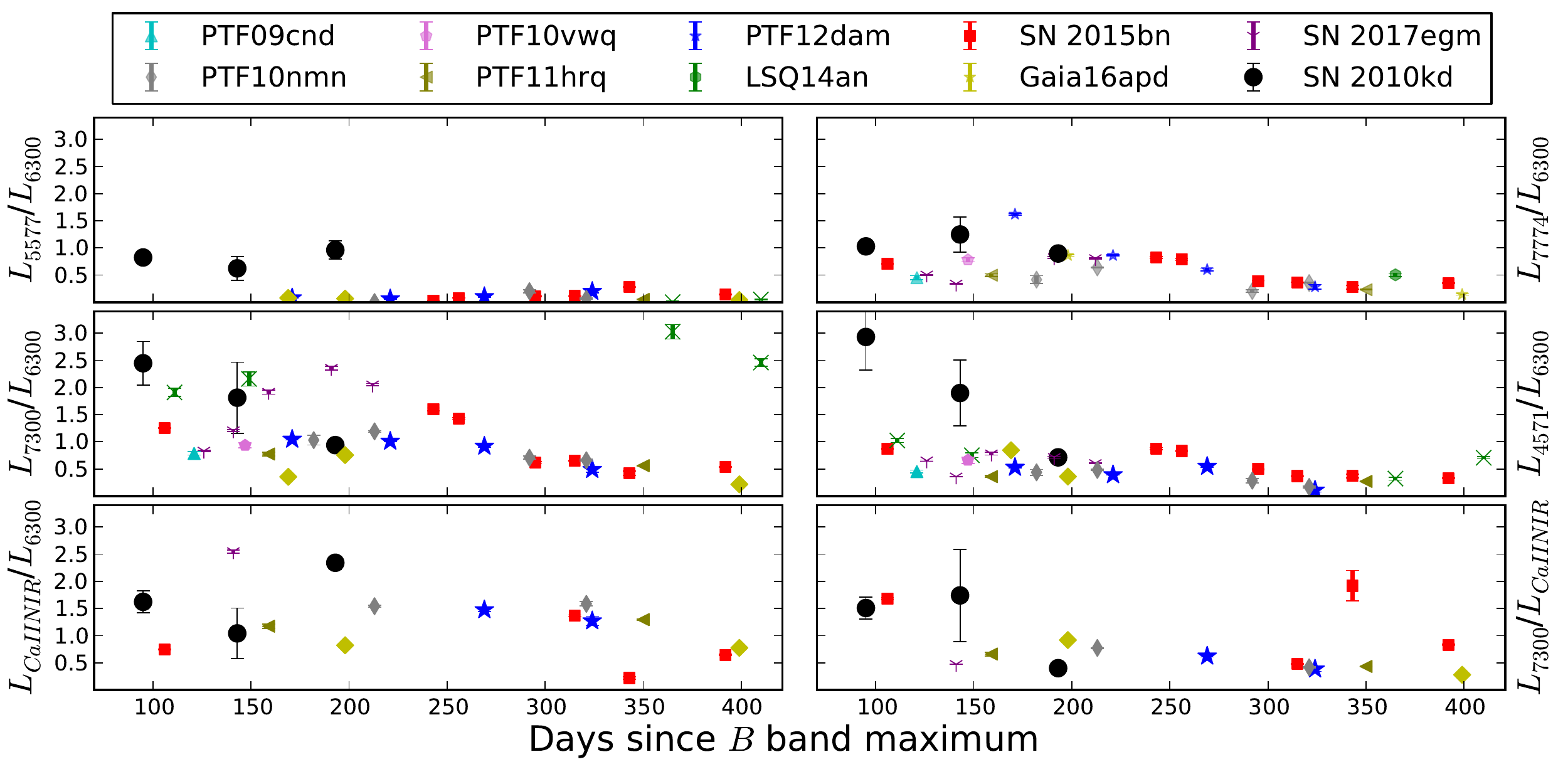}
\caption{Line-luminosity ratios of SN 2010kd are compared with well-studied SLSNe~I. [O~I] $\lambda$5577, O~I $\lambda$7774, Mg~I], [Ca~II], and the Ca~II NIR triplet luminosities are normalized to the luminosity of [O~I] $\lambda\lambda$6300, 6364. The [Ca~II] line-luminosity is normalized also with the Ca~II NIR triplet. For SN 2010kd, ${L_{5577}}/{L_{6300}}$ is higher than other presented SLSNe~I, which indicates a higher single-zone LTE temperature ($\sim 6000$~K).}
\label{fig:figlumcomprat}
\epsscale{9.}
\end{figure*}

In the case of SN 2010kd, the luminosity of the strongest lines is $\sim 10^{42}$ erg s$^{-1}$ at +96 days, dropping to $\sim 4 \times 10^{40}$ erg s$^{-1}$ at +194 days, consistent with the result of \cite{Nicholl2019} for different SLSNe~I. For SN 2010kd at early epochs, Mg~I] has the highest luminosity, followed by [Ca~II], the Ca~II NIR triplet, and O~I $\lambda$7774; the rest of the lines have even lower luminosities.

We compare luminosities of various lines of SN 2010kd with a set of well-studied SLSNe~I (see Figure~\ref{fig:figlumcomp}). All of the published comparison data were taken from \cite{Nicholl2019} and references therein; \cite{Nicholl2013,Nicholl2016a,Nicholl2016b,Nicholl2017,Chen2015,Inserra2017,Jerkstrand2017,Kangas2017,Quimby2018}. In the present study, for comparison we mainly discuss the line-luminosity evolution of PTF12dam and SN 2015bn, because these SLSNe~I have a significant number of spectroscopic data points and are also at comparable redshifts. Line-luminosity values of these objects are fitted with straight lines.

Throughout this section, the line-luminosity is in units of $10^{41}$ erg s$^{-1}$, whereas the decay rate is in $10^{41}$ erg s$^{-1}$ every 100 days. Line luminosities are compared at early phases ($\sim +96$ days), because at later phases, the luminosity values seem to converge. The [O~I] $\lambda\lambda$6300, 6364 line-luminosity is lower for SN 2010kd ($\sim 3.85$) in comparison to PTF12dam ($\sim 7.20$) and SN 2015bn ($\sim 5.75$), whereas the decay rate is nearly the same for SN 2010kd and PTF12dam ($\sim 2.80$). The O~I $\lambda$7774 line-luminosity of PTF12dam is higher ($\sim 1.10$), with a steeper decay rate ($\sim 5.15$) relative to SN 2010kd and SN 2015bn. SN 2010kd shows higher line luminosities of the [Ca~II] and Mg~I] lines, and steeper decay rates in comparison with other SLSNe~I. In the case of [O~I] $\lambda$5577, PTF12dam and SN 2015bn have almost constant decay rates ($\sim 0.70$) in comparison to the decay rate ($\sim 1.90$) for SN 2010kd.

In Figure~\ref{fig:figlumcomprat}, luminosity ratios of key diagnostic lines are plotted. [O~I] $\lambda$5577, O~I $\lambda$7774, Mg~I], [Ca~II], and the Ca~II NIR triplet are normalized to the luminosity of the [O~I] $\lambda\lambda$6300, 6364 doublet. The [Ca~II] line-luminosity is normalized also with the Ca~II NIR triplet luminosity. Line-luminosity ratios show that the luminosities of all presented lines are $\sim 1$--3 times the [O~I] $\lambda\lambda$6300, 6364 luminosity, except for [O~I] $\lambda$5577/[O~I] $\lambda\lambda$6300, 6364. Luminosity ratios at early phases (at +96 days) appear to have a larger range of values but converge at later epochs (+194 days).
Like other slow-decaying SLSNe~I, the [O~I] doublet in SN 2010kd appears to have less luminosity in comparison to [Ca~II] $\lambda7300$ and evolve much later, which indicate a central engine power source heating the inner ejecta as suggested by \cite{Nicholl2019}.

We also compared the line-luminosity ratios of various lines of SN 2010kd with other SLSNe~I. SN 2010kd has approximately the same range of values as other SLSNe~I for ${L_{\rm 7774}}/{L_{6300,~6364}}$, ${L_{\rm 7300}}/{L_{6300,~6364}}$, ${L_{\rm Ca~II~NIR}}/{L_{6300,~6364}}$, and ${L_{\rm 7300}}/{L_{\rm Ca~II~NIR}}$. In comparison to PTF12dam and SN 2015bn, SN 2010kd has a high ${L_{5577}}/{L_{6300,~6364}}$ luminosity ratio with a nearly similar decay rate, and high ${L_{\rm 4571}}/{L_{6300,~6364}}$ with sharper decay rate.

We calculated some physical parameters of SN 2010kd using these luminosity ratios, as done by \cite{Jerkstrand2014} and \cite{Nicholl2019} for other SLSNe~I. For SN 2010kd, ${L_{5577}}/{L_{6300,~6364}}$ is $\sim 0.5$--1.0, which is higher than those of other SLSNe~I: $\sim 0.1$--0.2 (see Figure~\ref{fig:figlumcomprat}). We assume that [O~I] $\lambda$5577 and the [O~I] doublet are from the same zone, so their ratio will depend only on optical depth and temperature. Thus, the ratio of these lines can indicate the single-zone local thermodynamic equilibrium (LTE) temperature, as discussed by \cite{Jerkstrand2017}, \cite{Nicholl2019}. To calculate the temperature, we used Equation~2 from \cite{Jerkstrand2014} and assume $\beta_{\rm ratio} = 1.5$. The temperature estimated for various SLSNe~I using nebular-phase spectra after +200 days is $\lesssim 5000$~K \citep{Jerkstrand2014,Nicholl2019}. But SN 2010kd has a very high single-zone LTE temperature ($\sim 6000$~K), estimated using spectra at +96, +144, and +194 days (see Table~\ref{tab:table7}). In summary, SN 2010kd has a higher ${L_{5577}}/{L_{6300,~6364}}$ luminosity ratio as well as a high single-zone LTE temperature in comparison to the other SLSNe~I presented in this study.

\begin{table}[]
  \begin{center}
    \caption{${L_{5577}}/{L_{6300,~6364}}$ and LTE temperature.\tnote{a}}
    \label{tab:table7}
    \begin{tabular}{c c c c} 

    \hline \hline

     Time & $\frac{L_{5577}}{L_{6300,~6364}}$ & LTE $T$ \\

      (days) & ($\beta_{\rm ratio}$ = 1.5) & (K) & $ $\\
    \hline

      +96 & 0.80 $\pm$ 0.10 & 6090 $\pm$ 190 &\\

      +144 & 0.60 $\pm$ 0.25 & 5700 $\pm$ 440 & \\

      +194 & 0.95 $\pm$ 0.30 & 6320 $\pm$ 520 &\\ 

      \hline 
    \end{tabular}

    \begin{tablenotes}[para,flushleft]
    \item[a] The line-luminosity ratio ${L_{5577}}/{L_{6300,~6364}}$ and 
the single-zone \\ LTE temperature corresponding to this ratio. For all of the \\ 
line-luminosity measurements, we choose $\beta_{\rm ratio} = 1.5$ as estimated\\ 
by \cite{Jerkstrand2014}.
    \end{tablenotes}

  \end{center}

\end{table}
 
In stripped-envelope SNe, the oxygen ejecta mass is $\sim70$\% of the total $M_{\rm ej}$ \citep{Maurer2010}. We can estimate the O~I ejected mass using the luminosity ratio ${L_{6300}}/{L_{6364}}$ \citep{Jerkstrand2014}. Unfortunately, our SN 2010kd spectra have low signal-to-noise ratio, so we are unable to measure ${L_{6300}}/{L_{6364}}$. However, we can estimate the upper limit of the O~I ejected mass by comparing the luminosity of the [O~I] $\lambda\lambda$6300, 6364 blend. \cite{Jerkstrand2017} explained that the spectra of SN 2007bi, LSQ-14an, and SN 2015bn can be reproduced only by a model having oxygen mass $\gtrsim 10~M_\odot$. \cite{Nicholl2019} compared the [O~I] $\lambda\lambda$6300, 6364 line-luminosity of many SLSNe~I (plotted in Figure~\ref{fig:figlumcomprat}) including SN 2007bi, LSQ-14an, and SN 2015bn and gave an upper limit of O~I ejected mass $\sim 10~M_\odot$. Similarly, we compared the [O~I] $\lambda\lambda$6300, 6364 line-luminosity of SN 2010kd with a set of well-studied SLSNe~I; see the upper-left panel of Figure~\ref{fig:figlumcomp}. We can see clearly that the [O~I] $\lambda\lambda$6300, 6364 line-luminosity of SN 2010kd is lower than that of SN 2007bi, PTF12dam, LSQ-14an, and SN 2015bn, and higher than that of SN 2017egm. So, for SN 2010kd, the upper limit of O~I ejected mass is $\sim 10~M_\odot$. This result is in good agreement with the total $M_{\rm ej}$ value calculated using the {\tt Minim} fitting to the light-curve of SN 2010kd (see Section~\ref{sec:MINIM}).

\subsection{Study of [O~III] Lines in the Nebular-phase of SN 2010kd}\label{sec:sec7}
\begin{figure}[ht!]
\includegraphics[angle=0,scale=1.1]{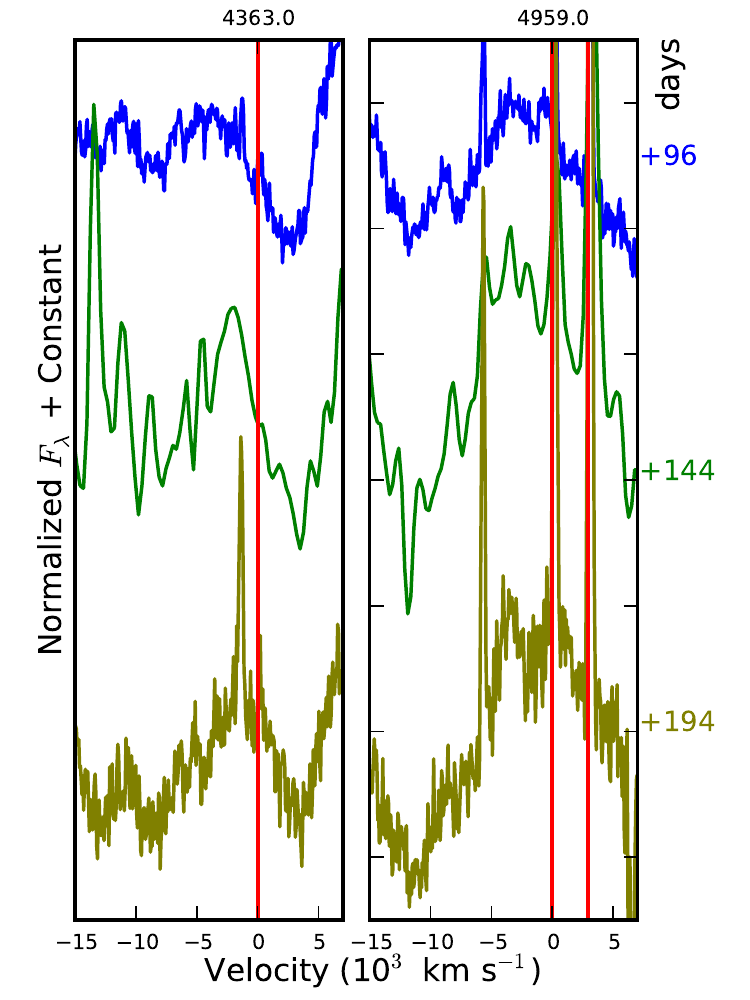}
\caption{The evolution of the broad [O~III] emission line of SN 2010kd at +96, +144, and +194 days. Left: evolution of [O~III] $\lambda$4363; zero velocity is set at 4363~\AA. Right: evolution of [O~III] $\lambda\lambda$4959, 5007; zero velocity is set at 4959~\AA. At +96 days, the [O~III] lines of SN 2010kd show ${f_{4959,~5007}}/{f_{4363}} \approx 1.6$; at $T \approx 8000$~K; this corresponds to an electron density of $\sim 6.0 \times 10^7$ cm$^{-3}$.}
\label{fig:figOIII}
\epsscale{6.}
\end{figure}

The nebular spectra of SN 2010kd show broad emission features around 4360--5000~\AA; we identify them as broad components of the [O~III] $\lambda$4363 and [O~III] $\lambda\lambda$4959, 5007 lines (see Figure~\ref{fig:figOIII}) as noticed in case of PS1-14bj by \cite{Lunnan2016}. We did not calculate their velocity because of blending with host-galaxy emission lines. We measured the flux ratio ${f_{4959,~5007}}/{f_{4363}}$ for three nebular spectra of SN 2010kd, which could provide information about the electron density of the emitting region \citep{Osterbrock2006}.

Fluxes were calculated using a single-Gaussian fit to [O~III] $\lambda$4363 and a double Gaussian fit to [O~III] $\lambda\lambda$4959, 5007. We estimated the values of ${f_{4959,~5007}}/{f_{4363}} \approx 1.6$, 3.8, and 3.7 using the nebular-phase spectra at +96, +144, and +194 days, respectively. For PS1-14bj, \cite{Lunnan2016} reported this ratio to be $\lesssim 3$ and claimed that the electron density in the [O~III]-emitting region is close to the critical density for these transitions ($>10^6$ cm$^{-3}$). For slow-evolving LSQ14an, \cite{Inserra2017} claimed ${f_{4959, 5007}}/{f_{4363}} \approx 1.8$, nearly equal to the value for SN 2010kd at +96 days. Using the spectrum of SN 2010kd at +96 days, we found a temperature $T \approx 8000$~K, corresponding to an electron density of $\sim 6.0 \times 10^7$ cm$^{-3}$ for the medium. This electron density is estimated using the formula from \cite{Osterbrock2006} or Equation~1 of \cite{Inserra2017}.

As shown in Figure~\ref{fig:figOIII}, the peak of the [O~III] $\lambda$4363 line is blueshifted relative to the rest-frame wavelength. Similarly, the [O~III] $\lambda\lambda$4959, 5007 lines have centroids that appear blueshifted in comparison to the rest-frame wavelength, showing velocities of $\sim 3000$ km s$^{-1}$.

\section{Host Galaxy of SN 2010kd}\label{sec:host}
Photometric and spectral studies of SLSN host galaxies provide enormous information about the nature of the environment and possible progenitors. On the basis of host-galaxy studies, \cite{Neill2011} were the first to discuss that SLSNe tend to occur in low-mass galaxies with high specific star-formation rates (specific SFRs). Thereafter, \cite{Chen2013} found the host of SN 2010gx to be one of the lowest metallicity dwarf galaxies. On the basis of a systematic sample study of 30 SLSN hosts, \cite{Lunnan2014} showed that the host galaxies of SLSNe~I are dwarfs similar to GRB hosts, with low luminosities. Later, many authors studied samples of known hosts of SLSNe~I and found that they have lower metallicities and higher SFRs \citep{Leloudas2015,Angus2017,Perley2016,Chen2017b,Schulze2018}.

In the present study, the host galaxy of SN 2010kd observed by {\it Swift}-UVOT in the \textit{u}-band at an apparent magnitude of $21.54 \pm 0.23$ (with a $4.7\sigma$ significance level; see 
Table~\ref{tab:host} in Appendix) is used to constrain the SFR. As discussed by \cite{Lunnan2014} and \cite{Angus2017}, the SFR of the host of SN 2010kd can be constrained using the rest-frame UV flux with the formula
\begin{equation}
{\rm SFR} = 1.4 \times 10^{-28} \times L_u {\rm (erg}~{\rm s}^{-1}~{\rm Hz}^{-1})
\end{equation} \citep{Kennicutt1998}. The derived value of the SFR of the host of SN 2010kd is $\sim 0.18 \pm 0.04~M_\odot$ yr$^{-1}$, which is in good agreement with the SFR calculated by \cite{Lunnan2014}, \cite{Leloudas2015}, and \cite{Schulze2018} and also within the range of SFRs estimated for other well-studied SLSN hosts.

\begin{table}[]
\scriptsize
  \begin{center}
    \caption{List of Host-galaxy Emission Lines of SN 2010kd.}
    \label{tab:EWfluxhost}
    \addtolength{\tabcolsep}{9pt}
    \begin{tabular}{c c c } 

    \hline \hline

    Line & Flux & EW (lower limit) \\

    $ $ & ($10^{-16}$ erg s$^{-1}$ cm$^{-2}$) & (\AA)  \\

    \hline

     H$\alpha$ & 5.67 $\pm$ 0.24  & 16.20 $\pm$ 0.71 \\
     
     H$\beta$  & 1.86 $\pm$ 0.27  & 3.9 $\pm$ 0.56\\
             
     H$\gamma$ & 1.02 $\pm$ 0.25  & 1.91 $\pm$ 0.47\\
            
     H$\delta$ & 0.45 $\pm$ 0.25  & 0.85$\pm$ 0.48\\
            
     [N~II] $\lambda$6584 & 0.06$\pm$ 0.4 & 0.19 $\pm$ 1.28\\
            
     [O~II] $\lambda$3727 & 2.39$\pm$ 0.45  & 4.54$\pm$ 0.86\\
     
     [O~III]$\lambda$4363 & 0.17$\pm$ 0.15  & 0.33$\pm$ 0.28\\
            
     [O~III]$\lambda$4959 & 2.97 $\pm$ 0.26  & 5.48 $\pm$0.48\\
        
     [O~III]$\lambda$5007 & 8.87 $\pm$ 0.36  & 17.78$\pm$ 0.73\\ 

    \hline 
    \end{tabular}

  \end{center}

\end{table}

We also traced H$\alpha$, H$\beta$, H$\gamma$, H$\delta$, [N~II] $\lambda$6584, [O~II] $\lambda\lambda$3726, 3729, [O~III] $\lambda$4363, and [O~III] $\lambda\lambda$4959, 5007 narrow host-galaxy emission lines in 
our nebular-phase spectra, particularly at +194 days (see Figure~\ref{fig:fignebphase}; narrow black arrows). These lines are similar to those observed by \cite{Lunnan2014} in late-time 
spectra of many SLSN hosts. In the case of SLSN iPTF13ehe, broad H$\alpha$ and H$\beta$ emission lines overlapping with narrow ones have been explained in terms of H-rich CSM at larger radii during late nebular-phases \citep{Yan2015,Yan2017b}. The absence of broad H$\alpha$ and H$\beta$ emission lines in the nebular-phase spectra of SN 2010kd at +194 days excludes the possibility of late-time interaction of the SN envelope with H-rich material. 

Observed narrow emission lines from the host are contaminated by the SN contribution itself; thus, their fluxes and lower limits of their equivalent widths (EWs) were estimated using a single-Gaussian fit (see Table~\ref{tab:EWfluxhost}). 
The measured flux values are generally higher than those of \cite{Lunnan2014}, perhaps owing
to a larger slit width or worse seeing. As discussed, in the present study we estimated the lower limit of EW[O~III] $\sim 17.85$ \AA. However, \cite{Leloudas2015} estimated the upper limit of EW[O~III] $\sim 190$ \AA, which placed the host galaxy of SN 2010kd in the category of extreme emission-line galaxies. The derived ratios of [O~III]/H${\beta}$ and [N~II]/H${\alpha}$ using the line fluxes (see Table~\ref{tab:EWfluxhost}) indicate that the host galaxy of SN 2010kd could be a low-luminosity, high-SFR dwarf galaxy having extreme emission lines as concluded by \cite{Leloudas2015}.

\section{Summary of Results}\label{sec:CONclusion}
We have presented a detailed study of the photometric and the spectroscopic evolution of SN 2010kd, one of the closest slow-decaying SLSNe~I. Using ROTSE-IIIb discovery data and other multiband data obtained using {\it Swift}-UVOT and the literature, we constructed a densely sampled optical light-curve. Spectroscopy is presented from $-$28 to +194 days, obtained using the HET-9.2~m and Keck-10~m telescopes. The temporal and spectral evolution of SN 2010kd is found similar to those of slow-decaying SLSNe~I: SN 2007bi, PTF12dam, and SN 2015bn. A summary of the other major findings of our present analysis is as follows:

(1) The \textit{B}-band light-curve of SN 2010kd has a peak luminosity close to that of PTF12dam and Gaia16apd, but higher in comparison to fast-decaying SLSNe~I at comparable redshifts. Its postpeak decay rate is similar to that of SN 2007bi, PTF12dam, and SN 2015bn (slow-decliners), and shallower than those exhibited by PTF10hgi, SN 2011ke, and SSS120810:231802-560926 (fast-decliners).

(2) The $(U-B)$ color evolution of SN 2010kd shows that at phases from $\sim +15$ to +90 days, it is brighter in the \textit{U}-band and gets redder faster in the \textit{U} than in the \textit{B}, \textit{V}, \textit{R}, and \textit{I} bands. This may be because of the cooling due to expansion or enhancement of the metallic lines in the UV or a combination of these two.

(3) Analytic light-curve model fitting to the bolometric light-curve of SN 2010kd using the code {\tt Minim} suggests that CSMI/MAG may be possible powering sources for SN 2010kd. 

(4) The SED of the SN 2010kd matches a hot BB in the optical bands. In the photospheric-phase, the continuum temperature of SN 2010kd looks somewhat hotter in comparison with PTF12dam and SN 2015bn. However, SN 2010kd seems to have a smaller BB radius in comparison to SN 2015bn.

(5) During the photospheric-phase, the spectra of SN 2010kd are dominated by the O~I, O~II, C~II, C~IV, and Si~II lines; however, they are overcome by heavy metallic lines in the cool photospheric-phase. Low- and high-velocity components of O~II and Si~II lines in the postmaximum spectra suggest possible nuclear burning fronts in the outer layers of the ejecta.

(6) Early and late-time spectral comparisons of SN 2010kd reveal that it is similar to SN 2007bi, PTF12dam, and SN 2015bn with minor discrepancies. At $\sim +96$ days, SN 2010kd has spectra similar to those of fast-decaying SLSNe~I at $\sim +35$ days, indicating a comparatively higher ejected mass of SN 2010kd.

(7) A systematic comparison of the late-time spectra of SN 2010kd with those of broad-lined SNe~Ic and those associated with LGRBs reveals that they have similar lines with different intensities, possibly providing clues to the variations in physical conditions among these events.

(8) SN 2010kd exhibit relatively flatter line-velocity curves, similar to those of SN 2015bn, but at comparatively higher values. The velocity of the Fe~II $\lambda5169$ line at $\sim +10$ days and its slow-decay rate up to $\sim +30$ days placed SN 2010kd in the category of slow-decliners. The flatter line-velocity curves of SN 2010kd indicates towards a central engine power source.

(9) In the nebular-phase spectra of SN 2010kd, the Mg~I] emission line has the highest luminosity of $\sim 10^{42}$ erg s$^{-1}$ at +96 days, dropping to $\sim 10^{40}$ erg s$^{-1}$ at +194 days. All of the emission lines show an overall decay in luminosity as phase increases.

(10) In comparison to other SLSNe~I, SN 2010kd shows a higher luminosity ratio of [O~I] $\lambda$5577 to [O~I] $\lambda\lambda$6300, 6364 emission lines, indicating a higher single-zone LTE temperature of SN 2010kd ($\sim 6000$~K).
However, the lower line-luminosity of [O~I] $\lambda\lambda$6300, 6364 in comparison to [Ca~II] $\lambda$7300 indicates a central engine power source. Comparison of the [O~I] $\lambda\lambda$6300, 6364 line-luminosity with that of other SLSNe~I suggests an upper limit of O~I ejected mass of $\sim 10~M_\odot$ for SN 2010kd.

(11) The nebular spectra of SN 2010kd show a broad component of the [O~III] $\lambda$4363 and $\lambda\lambda$4959, 5007 lines that suggests the electron density of the emitting medium to be $\sim 6.0 \times 10^7$ cm$^{-3}$ . These features are comparatively weaker in the spectra of PTF12dam.

(12) The SFR ($\sim 0.18 \pm 0.04~M_\odot$ yr$^{-1}$) for the host galaxy of SN 2010kd is constrained using the rest-frame UV flux. The derived flux ratios of [O~III]/H$\beta$ and [N~II]/H$\alpha$ are in agreement with previous findings, suggesting that the host of SN 2010kd is a low-luminosity, high-SFR dwarf galaxy.

\acknowledgments
This paper includes data taken from the RSVP program of the ROTSE-IIIb telescope (supported by NASA grant NNX10A196H, PI Carl Akerlof) at the McDonald Observatory of The University of Texas at Austin, 9.2m HET, and 10.0m Keck telescopes. 

The HET is a joint project of the University of Texas at Austin, the Pennsylvania State University, 
Stanford University, Ludwig-Maximilians-Universit\"{a}t M\"{u}nchen, and Georg-August-Universit\"{a}t G\"{o}ttingen. Some of the data presented herein were obtained at the W. M. Keck Observatory, which is operated as a scientific partnership among the 
California Institute of Technology, the University of California, and NASA.
This work also uses data supplied by the UK {\it Swift} Science Data Centre at the University of Leicester and {\it Swift}-UVOT data released 
in the {\it Swift} Optical/Ultraviolet Supernova Archive (SOUSA). 
S.B.P., K.M., D.B., A.A., and R.G. acknowledge BRICS grant DST/IMRCD/\\BRICS/Pilotcall/ProFCheap/2017(G) and DST/JSPS grant DST/INT/JSPS/P/281/2018 for this work. J.V. and R.K-T. are supported  by the  project  ``Transient  Astrophysical Objects'' GINOP 2.3.2-15-2016-00033 of the National Research, Development, and Innovation Office (NKFIH), Hungary, funded by the European Union. The research of J.C.W. is supported in part by NSF AST-1813825. A.V.F.'s supernova group is grateful for the support of the TABASGO Foundation, the Christopher Redlich Fund, and the Miller Institute for Basic Research in Science (U.C. Berkeley).
A.K. is grateful to Dr. Matt Nicholl for sharing the ASCII file from the published data. A. K. also acknowledges the PhD thesis of Rupak Roy as reference for some of the published photometric data taken with the 1.04m Sampurnanand Telescope of ARIES Nainital. A.K. and team members are grateful to the referee for his/her constructive comments that improved the manuscript.
This research has utilized the NED, which is operated by the Jet Propulsion Laboratory, California Institute of Technology, under contract with NASA. We acknowledge the availability of NASA ADS services. This research also made use of the Open Supernova Catalog (OSC) currently maintained by James Guillochon and Jerod Parrent.

\vspace{5mm}
\facilities{HET-9.2m, Keck:I, {\it Swift}.}      
          
\software{$Python$, matplotlib \citep{Hunter2007}, HEASOFT \citep[v6.26,][]{HEASARC2014}, $IRAF$ \citep{Tody1986,Tody1993}, {\it ximage} (v4.5.1), {\tt sms} \citep{Inserra2018c}, {\tt Superbol} \citep{Nicholl2018}, {\tt Minim} \citep{Chatzopoulos2013}, {\tt SYNAPPS} \citep{Thomas2011}}

\appendix

\subsection{Optical, {\it Swift}-UVOT, and {\it Swift}-XRT Data Reduction of SN 2010kd} \label{sec:Bband}

\subsubsection{Optical data analysis}
Reduction of ROTSE-IIIb data is discussed in Section~\ref{sec:ROTSElum}.
Published optical photometry in the Johnson $UBV$ and Cousins $RI$ bands is taken from \cite{Roy2012}.
The $B$- and $R$-band photometry also estimated from spectra using the code {\tt sms} \citep{Inserra2018c}.

\subsubsection{{\it Swift}-UVOT data analysis}
We analyzed all of the publicly available early-time (Target ID 00031890, 2010) and late-time (Target ID 00085565, 2014--2016) observations of {\it Swift}-UVOT.  As part of the present analysis, UVOT data were obtained from the online archive \footnote{http://swift.gsfc.nasa.gov/docs/swift/archive/} and analyzed using HEASOFT software v.~6.25 with the latest calibration database. We perform the reduction of the early-time UVOT data using the standard {\it uvotproduct} pipeline. A source circular region of $5''$ and background region of $50''$ radius were extracted for analysis. All of the estimated magnitudes in different filters are given in Tables~\ref{tab:my-table} and \ref{tab:host}.
 
UVOT again started following SN 2010kd during 2014--2016 in the $u$ filter.
In search of the host galaxy, all late-time sky images (Target ID 00085565) were stacked after their alignment to perform photometry of the source using the {\it uvotsource} pipeline (if visible after stacking or their upper limits). We summed the extensions within a sky image using {\it uvotimsum}. To sum the sky images from different observations, we merged the images first, using {\it fappend}. We considered the background threshold equal to three for calculating the magnitude of the source using {\it uvotsource}. A very faint extended object was detected, probably the host galaxy of SN 2010kd. Table~\ref{tab:host} shows the filter, start time $t_{\rm start}$, and stop time $t_{\rm stop}$ (relative to the trigger time), total exposure time, magnitude, and significance of the source in $u$ for the late-time observations. There is no correction for the Galactic extinction of the magnitudes given in Tables~\ref{tab:my-table} and \ref{tab:host}.

\subsubsection{{\it Swift}-XRT Data Analysis}
\label{XRTred}
The {\it Swift}-XRT also observed the field of SN 2010kd. We analyzed all of the early-time XRT observation at the position of SN 2010kd. For this purpose, XRT data were obtained from the online archive \footnote{https://www.swift.ac.uk/xrt$_{\textunderscore}$products/index.php} and analyzed using HEASOFT software v.~6.25 with the latest calibration database. We cleaned all of the XRT data using {\it xrtpipeline}, which removes the effects of hot pixels and Earth brightness. The {\it ximage} tool was used to detect the source, but no X-ray source was detected in individual images or in the stacked image near the UVOT position. 

We coadded all individual observations (Obs IDs) using the {\it ximage} software package (v.~4.5.1) to produce a sky image of SN 2010kd. We then used the {\it sosta} pipeline to determine the significance of the source's count-rate upper limit in individual Obs IDs as well as in the stacked image. These count rates were converted to an unabsorbed flux in Table~\ref{tab:xrt} using the online {\it pimms} tool (v.~4.9) using the appropriate column density for the host galaxy \citep{Kalberla2005} and adopting a power-law spectral model with photon index $\Gamma = 2$. We converted  unabsorbed flux into the luminosity using luminosity-distance information. We found that the unabsorbed flux and  X-ray luminosity of each Obs Id are typically similar to those estimated by \cite{Levan2013} and \cite{Margutti2018}.

\begin{table}[]
    \caption{{\it Swift}-UVOT Photometry of SN 2010kd.}
\scriptsize
  \begin{center}
    \label{tab:my-table}
    \addtolength{\tabcolsep}{35pt}
    \begin{tabular}{c c c c} 

    \hline \hline

    Filter & MJD & Magnitude & Obs ID \\
    \hline
    
\textit{u} & 55,530.19 & 16.46 $\pm$ 0.04 & 00031890001 \\

\textit{u} & 55,530.93 & 16.31 $\pm$ 0.07 & 00031890002 \\

\textit{u} & 55,546.31 & 16.11 $\pm$ 0.03 & 00031890003 \\

\textit{u} & 55,549.06 & 16.14 $\pm$ 0.03 & 00031890004 \\

\textit{u} & 55,552.39 & 16.22 $\pm$ 0.03 & 00031890005 \\

\textit{u} & 55,555.82 & 16.29 $\pm$ 0.03 & 00031890006 \\

\textit{b} & 55,530.19 & 17.47 $\pm$ 0.05 & 00031890001 \\

\textit{b} & 55,546.31 & 17.13 $\pm$ 0.03 & 00031890003 \\

\textit{b} & 55,549.06 & 17.15 $\pm$ 0.05 & 00031890004 \\

\textit{b} & 55,552.39 & 17.14 $\pm$ 0.03 & 00031890005 \\

\textit{b} & 55,555.82 & 17.20 $\pm$ 0.04 & 00031890006 \\

\textit{v} & 55,530.18 & 17.21 $\pm$ 0.07 & 00031890001 \\

\textit{v} & 55,530.91 & 17.11 $\pm$ 0.05 & 00031890002 \\

\textit{v} & 55,546.30 & 17.01 $\pm$ 0.05 & 00031890003 \\

\textit{v} & 55,549.05 & 17.07 $\pm$ 0.05 & 00031890004 \\

\textit{v} & 55,552.39 & 16.97 $\pm$ 0.04 & 00031890005 \\

\textit{v} & 55,555.81 & 17.01 $\pm$ 0.05 & 00031890006 \\

\textit{uvw1} & 55,530.18 & 17.35 $\pm$ 0.07 & 00031890001 \\

\textit{uvw1} & 55,530.92 & 16.97 $\pm$ 0.05 & 00031890002 \\

\textit{uvw1} & 55,546.30 & 16.81 $\pm$ 0.04 & 00031890003 \\

\textit{uvw1} & 55,549.06 & 16.90 $\pm$ 0.05 & 00031890004 \\

\textit{uvw1} & 55,552.39 & 16.99 $\pm$ 0.04 & 00031890005 \\

\textit{uvw1} & 55,555.82 & 17.11 $\pm$ 0.05 & 00031890006 \\

\textit{uvw2} & 55,530.18 & 18.69 $\pm$ 0.15 & 00031890001 \\

\textit{uvw2} & 55,530.91 & 18.30 $\pm$ 0.09 & 00031890002 \\

\textit{uvw2} & 55,546.30 & 17.95 $\pm$ 0.07 & 00031890003 \\

\textit{uvw2} & 55,549.05 & 17.96 $\pm$ 0.07 & 00031890004 \\

\textit{uvw2} & 55,552.39 & 18.13 $\pm$ 0.07 & 00031890005 \\

\textit{uvw2} & 55,555.81 & 18.22 $\pm$ 0.08 & 00031890006 \\

\textit{uvm2} & 55,530.18 & 18.14 $\pm$ 0.14 & 00031890001 \\

\textit{uvm2} & 55,530.92 & 17.65 $\pm$ 0.09 & 00031890002 \\

\textit{uvm2} & 55,546.30 & 17.51 $\pm$ 0.07 & 00031890003 \\

\textit{uvm2} & 55,549.05 & 17.47 $\pm$ 0.07 & 00031890004 \\

\textit{uvm2} & 55,552.39 & 17.59 $\pm$ 0.07 & 00031890005 \\

\textit{uvm2} & 55,555.81 & 17.70 $\pm$ 0.08 & 00031890006 \\

      \hline
    \end{tabular}

  \end{center}

\end{table}

\begin{table}[]
    \caption{Host-galaxy Magnitudes from Stacked Late-time {\it Swift}-UVOT Images}
\scriptsize
  \begin{center}
    \label{tab:host}
    \addtolength{\tabcolsep}{17pt}
    \begin{tabular}{c c c c c c c} 
    \hline       \hline
    
    Filter & $t_{\rm start}$(s) & $t_{\rm stop}$(s) & Exp. & Magnitude & Significance \\

    $ $ & MJD = 56,980.90 & MJD = 57,561.00 & (s) & (mag) & $ $ \\

          \hline
          
\textit{u} & 4.381254199412E+08 & 4.882461009676E+08 & 16,792.18 & 21.54 $\pm$ 0.23 & $4.7\sigma$ \\

      \hline
      
    \end{tabular}

  \end{center}

\end{table}

\begin{table*}[]
\begin{center}

\caption{Log of {\it Swift}-XRT Upper Limits on Unabsorbed Flux, 0.3--10 keV Band}
\label{tab:xrt}
\begin{tabular}{c c c c c c}
\hline
UT Date & MJD & Obs ID & \begin{tabular}[c]{@{}c@{}} Exposure\\ (s)\end{tabular} & \begin{tabular}[c]{@{}c@{}}Unabsorbed Flux (0.3--10 keV)\\ ($10^{-14}$ erg s$^{-1}$ cm$^{-2}$)\end{tabular} & \begin{tabular}[c]{@{}c@{}}$L_{X}$\\ ($10^{42}$ erg s$^{-1}$)\end{tabular} \\ \hline

2010 Nov 30 & 55530.13 & 00031890001 & 1785 & $< 23.96$ & $< 5.2$ \\

2010 Nov 30 & 55530.91 & 00031890002 & 1788 & $< 32.73$ & $< 8.6$ \\

2010 Dec 16 & 55546.29 & 00031890003 & 3181 & $< 13.49$ & $< 3.5$ \\

2010 Dec 19 & 55549.04 & 00031890004 & 2811 & $< 14.88$ & $< 3.9$ \\

2010 Dec 22 & 55552.38 & 00031890005 & 3476 & $< 12.17$ & $< 3.2$ \\

2010 Dec 25 & 55555.81 & 00031890006 & 3428 & $< 12.13$ & $< 3.2$ \\

 \hline
 
Stacked & Stacked &Stacked & 16,472 & $< 3.096$ & $< 0.8$ \\ \hline

\end{tabular}

\end{center}
\end{table*}

\newpage
\section{{\tt SYNAPPS} Spectral Matching}\label{sec:SYNAPPSapp}
The best-fit parameters obtained from {\tt SYNAPPS} spectral matching to the photospheric-phase spectra of SN 2010kd (see Figure~\ref{fig:figphotphase}) are listed in Table~\ref{tab:SYNAPPSapp}. These parameters are the line opacity $\tau$, lower and upper cutoff velocities $v_{\rm min}$ and $v_{\rm max}$ (respectively), e-folding length $aux$, and BB photospheric temperature $T$. It uses the $aux$ parameter for deciding the form of opacity and the Boltzmann excitation temperature for line-strength parameterization. The YAML \footnote{See http://www.yaml.org.} file consists of a few more parameters known as coefficients of the quadratic warping function, which is multiplied by the synthetic spectrum after proper computation. The limitations of this code are a sharp photosphere, BB assumption, and the approximation of no electron scattering.

\begin{longtable}{*6{p{2cm}}}
\caption{Best-fit parameters obtained using {\tt SYNAPPS} spectral matching.}\\
\hline
    Element & log$\tau$ & $v_{\rm min}$ & $v_{\rm max}$ & $aux$ & $T$ \\ 
       $ $     &  $ $ & ($10^3$ km s$^{-1}$) & ($10^3$ km s$^{-1}$) &  & ($10^3$ K)  \\ 
    \hline
    \hline
    
At $-28$ days  &		&		&		&		& \\
  
\hline

C~II   &  $-1.25$ & 12.0 & 41.14 & 12.0 & 15.32 \\

C~IV   &  $-0.54$ & 14.5 & 20.44 & 6.97 & 18.43\\

O~I    &   0.4 & 11.79 & 30.0 & 1.45 & 19.23   \\

O~II   &  $-0.28$ & 12.44 & 29.92 & 0.42 & 17.11 \\

Si~II  &   0.23 & 21.99 & 40.85 & 5.5 & 6.01  \\

Fe~III &  $-0.27$ & 11.8 & 30.87 & 1.58 & 16.72 \\

\hline

At $-24$ days	&		&		&		&		&	      \\

\hline

C~II   &  $-1.02$ & 11.94 & 41.53 & 4.99 & 14.31\\

C~IV   &  $-0.61$ & 14.02 & 22.37 & 6.33 & 25.0\\

O~I    &   0.42 & 11.69 & 30.0 & 1.4 & 19.03\\

O~II   &  $-0.5$ & 12.24 & 29.93 & 0.42 & 17.08\\

Si~II  &  $-0.66$ & 13.26 & 44.91 & 10.0 & 11.86\\

Fe~III &  $-1.51$ & 11.71 & 29.9 & 2.1 & 5.0\\

\hline

At $-23$ days	&		&		&		&		&	      \\

\hline

C~II   &  $-1.18$ & 11.85 & 40.59 & 5.69 & 13.48\\

C~IV   &  $-0.58$ & 14.07 & 24.12 & 6.17 & 25.0\\

O~I    &   0.4 & 11.63 & 30.0 & 1.38 & 18.83\\

O~II   &  $-1.36$ & 12.14 & 29.93 & 0.42 & 19.44\\

Si~II  &  $-0.75$ & 13.48 & 40.31 & 10.0 & 10.56\\

Fe~III &  $-0.36$ & 11.7 & 29.87 & 1.81 & 14.79\\

\hline

At $-22$ days	&		&		&		&		&	      \\

\hline

C~II   &  $-1.31$ & 11.89 & 34.58 & 10.0 & 14.26\\

C~IV   &  $-0.73$ & 13.89 & 21.25 & 10.0 & 22.99\\

O~I    &   0.39 & 11.61 & 30.0 & 1.43 & 18.23\\

O~II   &  $-0.53$ & 12.01 & 29.93 & 0.42 & 16.66\\

Si~II  &  $-0.29$ & 22.75 & 38.46 & 9.44 & 7.08\\

Fe~III &  $-1.25$ & 11.64 & 26.58 & 12.0 & 9.98\\

\hline

At +15 days	&		&		&		&		&	      \\

\hline

C~II  &  $-1.43$ & 11.35 & 49.42 & 3.29 & 11.03\\

O~I   &   0.17 & 11.39 & 49.97 & 1.96 & 5.6\\

O~II LV &  $-0.47$ & 12.02 & 49.93 & 3.15 & 11.1\\

O~II HV &  $-0.63$ & 21.5 & 49.91 & 14.99 & 11.1\\

Na~I  &  $-0.64$ & 11.35 & 49.89 & 5.47 & 19.88\\

Mg~II &  $-0.25$ & 11.92 & 49.92 & 2.83 & 13.23\\

Si~II LV &  $-0.03$ & 11.35 & 49.84 & 2.92 & 17.13\\

Si~II HV &  $-0.13$ & 19.0 & 49.97 & 8.36 & 17.13\\

Fe~II &  $-0.67$ & 11.94 & 49.37 & 5.64 & 8.22\\

\hline

At +34 days	&		&		&		&		&	      \\

\hline

C~II  &  $-1.92$ & 11.1 & 49.88 & 10.0 & 12.86\\

O~I   &   0.17 & 11.05 & 49.93 & 1.98 & 5.83\\

O~II LV &  $-0.63$ & 11.75 & 49.43 & 5.29 & 11.08\\

O~II HV &  $-0.64$ & 19.52 & 49.95 & 15.0 & 11.08\\

Na~I  &  $-0.37$ & 11.05 & 49.9 & 2.8 & 5.21\\

Mg~II &  $-0.34$ & 11.38 & 49.91 & 2.85 & 20.0\\

Si~II LV &  $-0.05$ & 11.05 & 49.92 & 2.57 & 17.49\\

Si~II HV &  $-0.11$ & 18.73 & 39.78 & 8.73 & 17.49\\

Fe~II &  $-0.64$ & 11.36 & 42.75 & 9.81 & 10.35\\

\hline

\label{tab:SYNAPPSapp}
\end{longtable}

\end{document}